\documentclass[preprint,12pt]{elsarticle}
\usepackage{color}
\usepackage[utf8]{inputenc}
\DeclareUnicodeCharacter{0301}{\'{e}}
\usepackage[margin=1.25in,footskip=0.5in]{geometry}

\usepackage[colorlinks=true, allcolors=green]{hyperref}
\hypersetup{colorlinks = true,linkcolor = blue,anchorcolor =red,citecolor = blue,filecolor = red,urlcolor = red, pdfauthor=author}

\usepackage{tabularx}
\usepackage{adjustbox}
\usepackage{subfig}
\usepackage{amsmath}
\usepackage{booktabs}
\usepackage{multirow}

\usepackage{amssymb}

\journal{Elsevier}

\begin{document}

\begin{frontmatter}



\title{DC-cycleGAN: Bidirectional CT-to-MR Synthesis from Unpaired Data}


\author[First]{Jiayuan Wang}
\ead{wang621@uwindsor.ca}
\author[First]{Q. M. Jonathan Wu\corref{cor1}}
\ead{jwu@uwindsor.ca}
\author[First]{Farhad Pourpanah}
\ead{farhad.086@gmail.com}

\cortext[cor1]{Corresponding author}

\address[First]{University of Windsor, 401 Sunset Ave, Windsor, N9B 3P4, ON, Canada}

\begin{abstract}
Magnetic resonance (MR) and computer tomography (CT) images are two typical types of medical images that provide mutually-complementary information for accurate clinical diagnosis and treatment. However, obtaining both images may be limited due to some considerations such as cost, radiation dose and modality missing.
Recently, medical image synthesis has aroused gaining research interest to cope with this limitation. In this paper, we propose a bidirectional learning model, denoted as \textit{dual contrast cycleGAN} (DC-cycleGAN), to synthesize medical images from unpaired data. 
Specifically, a \textit{dual contrast} loss is introduced into the discriminators to indirectly build constraints between real source and synthetic images by taking advantage of samples from the source domain as negative samples and enforce the synthetic images to fall far away from the source domain.
In addition, cross-entropy and structural similarity index (SSIM) are integrated into the DC-cycleGAN in order to consider both the luminance and structure of samples when synthesizing images. The experimental results indicate that DC-cycleGAN is able to produce promising results as compared with other cycleGAN-based medical image synthesis methods such as cycleGAN, RegGAN, DualGAN, and NiceGAN. The code will be available at https://github.com/JiayuanWang-JW/DC-cycleGAN. 
\end{abstract}




\end{frontmatter}


\section{Introduction}
\label{sec:Int}
Magnetic resonance (MR) and computed tomography (CT) images are of great importance and widely used for the diagnosis of various diseases such as cancer, as well as radiotherapy treatment planning~\cite{Yueyun2008ANALYSIS}. On one hand, MR images are excellent in capturing the diversity of contrasts and accurately locating tumors and organs for structural images with soft tissues~\cite{mastrogiacomo2019magnetic}. 
On the other hand, CT images are able to show details of the lesions and provide electron density information~\cite{wang2019mri}. Since, unique information can be obtained from each modality, using both MR and CT images of the patients can dramatically improve clinical diagnosis and treatment. Nonetheless, some considerations, including cost, time, radiation dose, and the necessity of accurate MR/CT registrations, may limit the acquisition of both images. These limitations can be avoided by synthesizing medical images from another modality, i.e., generating MR images from CT images~\cite{nie2018medical}.\par

Medical image synthesis can be defined as a mapping between target and source images. Existing methods can be grouped into feature mapping and deep learning (DL) methods. Feature mapping-based methods, which establish a mapping between similar patches of two modalities, can be further divided into atlas-, segmentation-, patch-, and sparse coding-based methods~\cite{xu2020bpgan}. 
As an example, atlas-based methods~\cite{yu2019ea}, first, measure the atlas-to-image transformation from source modality using the paired image atlases from both source and target modalities and then use this transformation for generating images in the target modality. Feature mapping methods usually need many pre-/post-processing steps as well as prior knowledge for tuning the transformation parameters. In addition, these methods face problems in generating low-resolution images, and they are not able to produce a variety of images with large anatomical differences~\cite{bi2017synthesis}.
In contrast, DL-based methods directly learn complex non-linear mappings from the source domain to the target domain~\cite{zhou2021survey,wang2020recent,pourpanah2020review}. 
Convolutional neural networks (CNNs)~\cite{krizhevsky2012imagenet} are among the most popular DL-based methods for medical image synthesis due to their ability in learning image structure. 
Nie at al.~\cite{nie2016estimating} synthesized CT images from MR images by adopting a 3D fully convolutional neural network (FCN). Fu et al.~\cite{fu2018male} used a U-Net structure to synthetic CT images from MR images.
Recently, generative-based methods, including generative adversarial networks (GANs)~\cite{goodfellow2014generative} and Variational autoencoders (VAEs)~\cite{kingma2013auto}, have shown remarkable results in synthesizing medical images of one modality conditioned on another modality~\cite{bi2017synthesis}.\par

\begin{figure}[thb!]
	\centering
	\subfloat[\small Real]{
		\begin{minipage}[t]{0.2\textwidth}
			\centering
			\includegraphics[width=\textwidth]{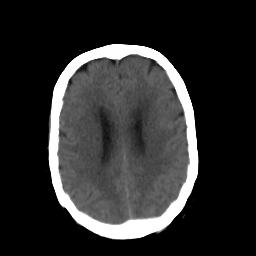}\\
			\vspace{0.1cm}
			\includegraphics[width=\textwidth]{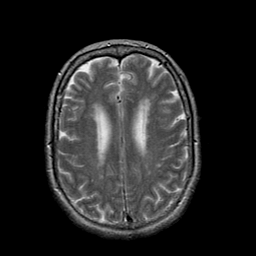}\\
			\vspace{0.35cm}
		\end{minipage}%
	}%
	\subfloat[\small MAE\&MSE]{
		\begin{minipage}[t]{0.2\textwidth}
			\centering
			\includegraphics[width=\textwidth]{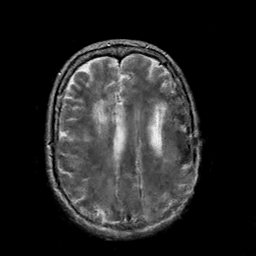}\\
			\vspace{0.1cm}
			\includegraphics[width=\textwidth]{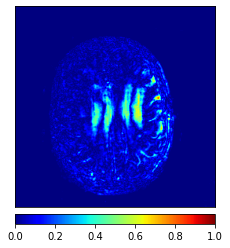}\\
			\vspace{0.1cm}
		\end{minipage}%
	}%
	\subfloat[\small SSIM\&CE]{
		\begin{minipage}[t]{0.2\textwidth}
			\centering
			\includegraphics[width=\textwidth]{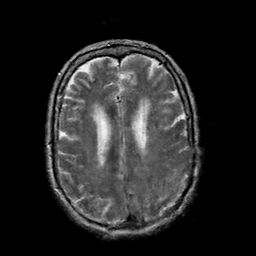}\\
			\vspace{0.1cm}
			\includegraphics[width=\textwidth]{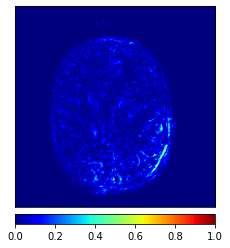}\\
			\vspace{0.1cm}
		\end{minipage}%
	}%
	\subfloat[\small Ours]{
		\begin{minipage}[t]{0.2\textwidth}
			\centering
			\includegraphics[width=\textwidth]{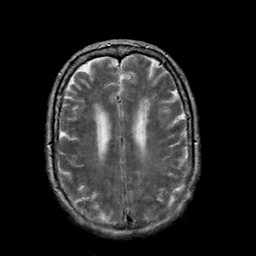}\\
			\vspace{0.1cm}
			\includegraphics[width=\textwidth]{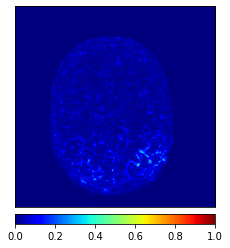}\\
			\vspace{0.1cm}
		\end{minipage}%
	}%

	\centering
	\caption{Synthesized MR images along with the absolute error maps between ground-truth and synthesized images by MAE \& MSE, CE \& SSIM without {\color{black}dual contrast loss} and CE \& SSIM with {\color{black}dual contrast losses (Ours)}. (a) shows the real images.}
	\label{MAEMSE_cessim_a2b}
\end{figure}

\begin{figure}[thb!]
	\centering
	\subfloat[\small Real]{
		\begin{minipage}[t]{0.2\textwidth}
			\centering
			\includegraphics[width=\textwidth]{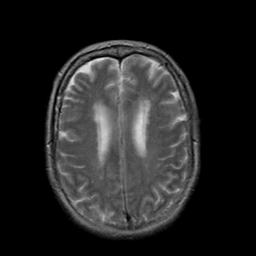}\\
			\vspace{0.1cm}
			\includegraphics[width=\textwidth]{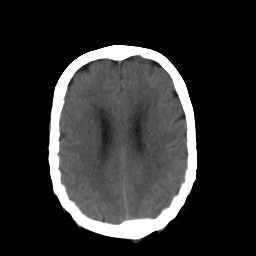}\\
			\vspace{0.35cm}
		\end{minipage}%
	}%
	\subfloat[\small MAE\&MSE]{
		\begin{minipage}[t]{0.2\textwidth}
			\centering
			\includegraphics[width=\textwidth]{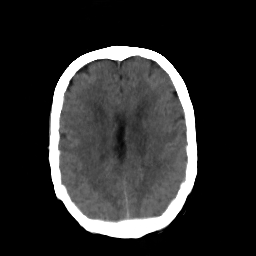}\\
			\vspace{0.1cm}
			\includegraphics[width=\textwidth]{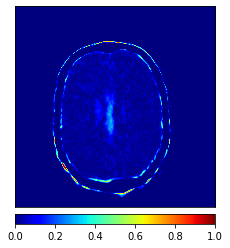}\\
			\vspace{0.1cm}
		\end{minipage}%
	}%
	\subfloat[\small SSIM\&CE]{
		\begin{minipage}[t]{0.2\textwidth}
			\centering
			\includegraphics[width=\textwidth]{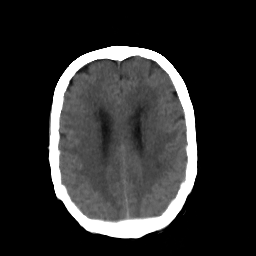}\\
			\vspace{0.1cm}
			\includegraphics[width=\textwidth]{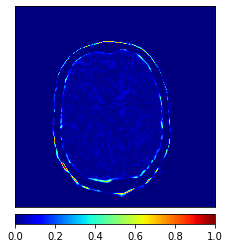}\\
			\vspace{0.1cm}
		\end{minipage}%
	}%
	\subfloat[\small Ours]{
		\begin{minipage}[t]{0.2\textwidth}
			\centering
			\includegraphics[width=\textwidth]{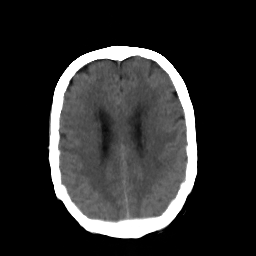}\\
			\vspace{0.1cm}
			\includegraphics[width=\textwidth]{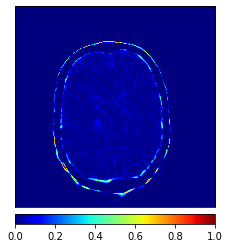}\\
			\vspace{0.1cm}
		\end{minipage}%
	}%

	\centering
	\caption{Synthesized CT images along with the absolute error maps between ground-truth and synthesized images by MAE \& MSE, CE \& SSIM without {\color{black}dual contrast loss} and CE \& SSIM with {\color{black}dual contrast losses (Ours)}. (a) shows the real images.}
	\label{MAEMSE_cessim_b2a}
\end{figure}

The main limitation of the above-mentioned methods is their dependency on a large number of paired images, i.e., both images belong to the same patient, perfectly registered for training~\cite{roy2014pet}, which is difficult to obtain. If the registration has a local mismatch between different modalities, the learning models would generate irrelevant images. 
To overcome this issue, Zhu et al.~\cite{zhu2017unpaired} proposed cycleGAN to synthesize images from unpaired data in an unsupervised manner.
Although cycleGAN has shown remarkable results in reconstructing identical images to the real input, e.g. CT-to-CT, it can not be directly used to synthesize samples from another modality, e.g., MR-to-CT, as there are no direct constraints between real source and synthetic images~\cite{Heran2020Unsupervised}.
For example, Zhang et al.~\cite{zhang2018translating} alleviated this issue using an additional loss to force the generated images to be the same as the real ones. Later, SC-cycleGAN~\cite{Heran2020Unsupervised} defined a structure-consistency loss into the cycleGAN. Specifically, modality-independent neighborhood descriptor (MIND)~\cite{heinrich2012mind} and a position-based selection strategy is used as structural features and slice selection, respectively. Moreover, several studies conducted bidirectional prediction, i.e., synthesis of CT images from given MR images and vice versa~\cite{xu2020bpgan,abu2021paired}.
In addition, many cycleGAN-based methods have been proposed for medical image synthesis from unpaired data \cite{hiasa2018cross,santini2020unpaired,chartsias2017adversarial}, but these methods only used the basic discriminator. The basic discriminator only utilizes images in the target domain, while we discovered that using reference images from only the target domain leads to a problem, as shown in Fig. \ref{MAEMSE_cessim_a2b} and \ref{MAEMSE_cessim_b2a}.\par

\par

In this paper, we propose a bidirectional learning model, known as \textit{dual contrast cycleGAN (DC-cycleGAN)}, for medical image synthesis from unpaired data.
Specifically, a dual contrast (DC) loss is formulated that leverages the advantage of samples from the source domain as negative samples to indirectly build constraints between real source and synthetic images via discriminators, and synthesize images more related to the target domain by enforcing the synthetic images to fall far away from the source domain. 
In addition, structural similarity index (SSIM)~\cite{Zhou2004Image} and cross-entropy (CE)~\cite{zhou2019mpce} are integrated into the {\color{black}DC-cycleGAN} structure to avoid disappearing gradient information that is caused by a mean absolute error (MAE) and synthesizing irrelevant images. SSIM considers luminance~\cite{Zhou2004Image} and CE converges fast as its back-propagation error is less than MSE~\cite{sangari2015convergence}. 
As can be seen in Figs.~1 and~2, using SSIM and CE with {\color{black}dual contrast} can generate more clear and accurate MR images as compared with that of MAE and MSE, and SSIM and CE without {\color{black}dual contrast} loss. Although both SSIM and CE with {\color{black}dual contrast} and without {\color{black}dual contrast} generate similar CT images, SSIM and CE with {\color{black}dual contrast} quantitatively generate better images as shown in Table~\ref{ablation_MRI-CT}.     
The experimental results indicate that {\color{black}DC-cycleGAN} is able to consider more complex features such as structure in synthesizing images and produce remarkable results as compared with other state-of-the-art methods reported in the literature.

The rest of the article is organized as follows. Section~\ref{sec:related2} presents the related work. The proposed bidirectional medical image synthesis model is presented in Section~\ref{Sec:meth}. The experimental results and discussions are provided in Section~\ref{sec:exp}. Finally, Section~\ref{Sec:con} concludes the paper and suggests future research directions.

\section{Related works}
\label{sec:related2}
Synthesizing CT/MR images from another modality can significantly reduce the cost and help to accurately diagnose various diseases. However, it is a challenging issue to synthesize one modality from another one as there are no direct constraints between MR and CT images. In general, medical image synthesize methods can be trained either from paired data, i.e., aligned CT and MR images from the same patient, or unpaired data. 
In the following subsections, we review these two categories.

\begin{figure}[tb!]
\centering
\includegraphics[width=0.990\linewidth]{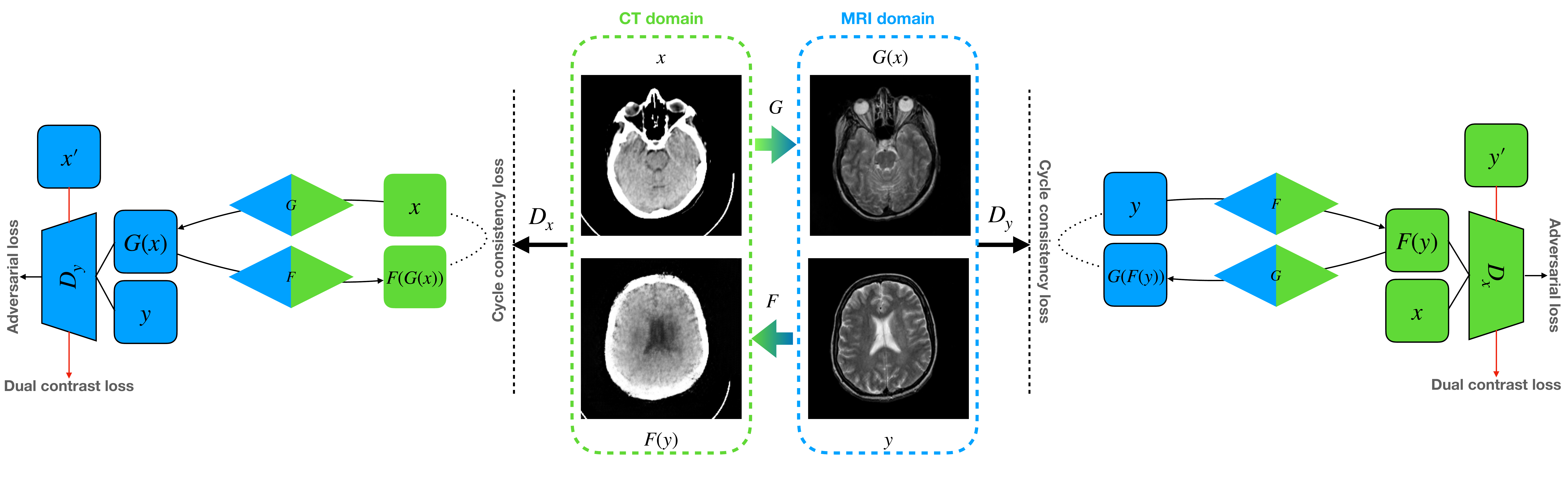}
\caption{The proposed {\color{black}DC-cycleGAN} model for medical image synthesize from unpaired data. It receives real CT/MR images through a generator to synthesize MR/CT images and discriminators distinguish real images from generated and real images from the source domain. Note that $x'$ and $y'$ are negative samples that are randomly selected from source images.}
\label{DC-cycleGAN}
\end{figure}

\subsection{Medical image synthesis using paired data }
\label{Sec:sec:med}

Most of the traditional medical image synthesis methods are usually trained based on paired data. For example, a method based on multi-atlas registration and label propagation approach has been proposed to synthesize CT images from MR images in~\cite{merida2015evaluation}. This method uses a pairwise registration technique to align the atlas MR and target images and then generates CT images through a voxelwise atlas selection and intensity averaging. 
The proposed segmentation-based method proposed in~\cite{hsu2013investigation} spatially aligns all MR volumes to the T1-weighted volume, then uses fuzzy c-means (FCM) clustering algorithm to compute the probability of belonging to a class for each voxel of images.  
Yang et al.~\cite{yang2018predicting} developed an approach that learns non-linear local descriptors and feature matching to generate CT images. This method projects linear descriptors into a nonlinear high-dimensional space to obtain non-linear descriptors. 
Then, it finds the nearest neighbors of the local descriptor in the MR images to estimate CT patches.  

Recently, with the development of digital technologies, DL-based medical image synthesis methods have received more attention. Chen et al.~\cite{chen2018u} adopted a 2D-U-Net to synthesize CT slices from T2-weighted MR images. Nie et al.~\cite{nie2017medical} integrated a 3D convolutional neural network (FCN) into the GAN structure to learn a mapping between CT and MR images.
RU-ACGAN~\cite{Qian2020Estimating} combines ResNet, U-net and ACGAN~\cite{odena2017conditional} for CT estimation. Despite the success of these methods in synthesizing medical images from another modality, they require a large number of paired images, which is difficult in practice.

\subsection{Medical image synthesis using unpaired data}
\label{Sec:sec:unsuper}
To solve the problem of learning from paired data, unsupervised learning techniques have been developed that are able to learn from unpaired data. 
FedMed~\cite{xie2022fedmed}, which is a self-supervised learning method, generates brain images from unpaired data. A data augmentation technique is introduced for self-supervised learning and then feds it into three auxiliary heads.
Sohail et al.~\cite{sohail2019unpaired} used GAN for multi-contrast MR image synthesis from unpaired data. Specifically, star-GAN is used to translate given image into a number of contrasts, and then introduced a new loss function for the generator to synthesize high-quality images. 
Chartsias et al.~\cite{chartsias2017adversarial} synthesized MR images from CT images to improve the segmentation of the cardiac dataset.   

Wolterink et al.~\cite{wolterink2017deep} used a 2D cycleGAN to synthesize images from unpaired images. 
However, using 2D cycleGAN for constructing 3D volume may lead to observing spatial consistency in the generated images.     
Later, Zeng et al.~\cite{zeng2019hybrid} used a 3D FCNs as a generator to better model the spatial information and alleviate the discontinuity problem across slices.
In~\cite{SeungKwan2021Synthetic}, a cycleGAN with perceptual loss is used to train a model with weakly paired CT and MR images. 
Pseudo-3D cycleGAN~\cite{oulbacha2020mri} integrates neighboring slices as well as cyclic loss function to ensure the consistency between MR and CT images.
UagGAN~\cite{abu2021paired} integrates an unsupervised attention mechanism into GAN to improve the context misalignment problem during learning from unpaired data. Other studies that added additional losses into cycleGAN structure to synthesize high-quality images from unpaired data including cycleGAN with a shape consistency loss~\cite{zhang2018translating}, cycleGAN with a gradient consistency loss~\cite{hiasa2018cross}, cycleGAN with adaptive instance normalization~\cite{yang2021continuous} and cycleGAN with structure-consistency loss \cite{Heran2020Unsupervised}, just to name a few. Inspired by the loss correction, RegGAN~\cite{kong2021breaking} trains a generator with an additional registration network to adaptively fit the misaligned noise distribution.

In this study, we propose a bidirectional learning model based on cycleGAN to synthesize MR/CT images from unpaired data. Compared with other methods, our method uses the advantages of samples from the source domain as negative samples during training to fall the synthesized images far away from the source domain and generate high-quality samples.

\section{Method}
\label{Sec:meth}
In this section, we discuss the proposed {\color{black}DC-cycleGAN} model in detail. First, an overview of {\color{black}DC-cycleGAN} is provided, and then each component is discussed.        

\subsection{Model overview}
\label{Sec:sec:model}
In this section, we discuss our proposed {\color{black}DC-cycleGAN} model for medical image synthesis in detail. 
The aim is to learn a bidirectional mapping function between MR and CT images using unpaired data, i.e., using MR and CT images from different subjects. Assume $X=\{x_i\}_{i=1}^{N}$ and $Y=\{y_j\}_{j=1}^{M}$ indicate sets of CT and MR images, respectively, where $x_i\in X$ and $y_i\in Y$.
Similar to the cycleGAN, our {\color{black}DC-cycleGAN} model (see Fig.~\ref{DC-cycleGAN}) consists of two generators $G:x\to \hat{y}$ and $F:y\to \hat{x}$ for learning CT-to-MR and MR-to-CT mappings, respectively, where $\hat{y}=G(x)$ and $\hat{x}=F(y)$ represent synthesized MR and CT images, {\color{black}and two discriminators $D_Y$ and $D_X$ for distinguishing real MR and CT images from the synthetic or negative ones, respectively.
In another word, the discriminators aim to distinguish whether the input image belongs to class 1, i.e., real image, or class 0, i.e., synthesized image or a sample from the source domain. In addition, each discriminator includes a {\color{black}DC loss} that leverage the advantages of samples from the source domain as negative samples to enforce the synthesized images falling far away from the source domain.}

\subsection{CycleGAN}
\label{Sec:CycleGAN}

GAN~\cite{goodfellow2014generative} consists of a generator $G$ and discriminator $D_Y$. The generator learns a mapping function $G: X\to Y$ to map the source domain $X$ to the target domain $Y$ 
, while the discriminator computes the probability that a sample $x$ belongs to the training data rather than synthesised one. $D_Y$ and $G$ are optimized by playing a minmax game, as follows:
\begin{equation}
\label{CycleGAN_adversarial_loss}
\mathcal{L}_{\mathrm{GAN}}\left(G, D_{Y}, X, Y\right) =\mathbb{E}_{y \sim p_{\mathrm{data}}(y)}\left[\log D_{Y}(y)\right] \\
+\mathbb{E}_{x \sim p_{\mathrm{data}}(x)}\left[\log \left(1-D_{Y}(G(x))\right)]\right..
\end{equation}


Since the original GAN learns a mapping to produce outputs identically distributed in the target domain, it can map a given input to any random space in the target domain. Thus, GAN's training loss can not guarantee to the production of the desired output for a given input. One possible solution for this issue is to reconstruct back the synthesized samples into the source domain using a cycle consistency loss~\cite{zhu2017unpaired}. This requires training a generator $F:Y\to X$ for reconstructing real samples and a discriminator $D_X$ for distinguishing real images from the generated ones. The cycle consistency loss can be written as:
\begin{equation}
\label{cycle_consistency_loss}
\mathcal{L}_{\mathrm{cycle}}(G, F) =\mathbb{E}_{x \sim p_{\text {data }}(x)}\left[\|F(G(x))-x\|_{1}\right] \\
+\mathbb{E}_{y \sim p_{\text {data }}(y)}\left[\|G(F(y))-y\|_{1}\right].
\end{equation}

The objective function for cycleGAN can be written as:
\begin{equation}
\mathcal{L}\left(G, F, D_{X}, D_{Y}\right)=\mathcal{L}_{\mathrm{GAN}}\left(G, D_{Y}, X, Y\right) \\
+\mathcal{L}_{\mathrm{GAN}}\left(F, D_{X}, Y, X\right)
+\lambda \mathcal{L}_{\mathrm{cycle}}(G, F),
\end{equation}
where $\lambda$ is the weight between the two-loss functions.

However, Directly applying cycleGAN to synthesize samples from another modality, i.e, MR-to-CT, cannot generate high-quality images as there are no direct constraints between real source and synthetic images~\cite{Heran2020Unsupervised}. In other words, using adversarial and cycle consistency losses cannot guide the generator to learn a robust mapping in the target domain. To address this issue, in this study, we leverage the advantage of samples from the source domain as negative samples during the training of discriminators. To achieve this, the concept of {\color{black}dual contrast} is used, which is discussed the next subsection.

\begin{figure}[tb!]
\centering
\includegraphics[width=\linewidth]{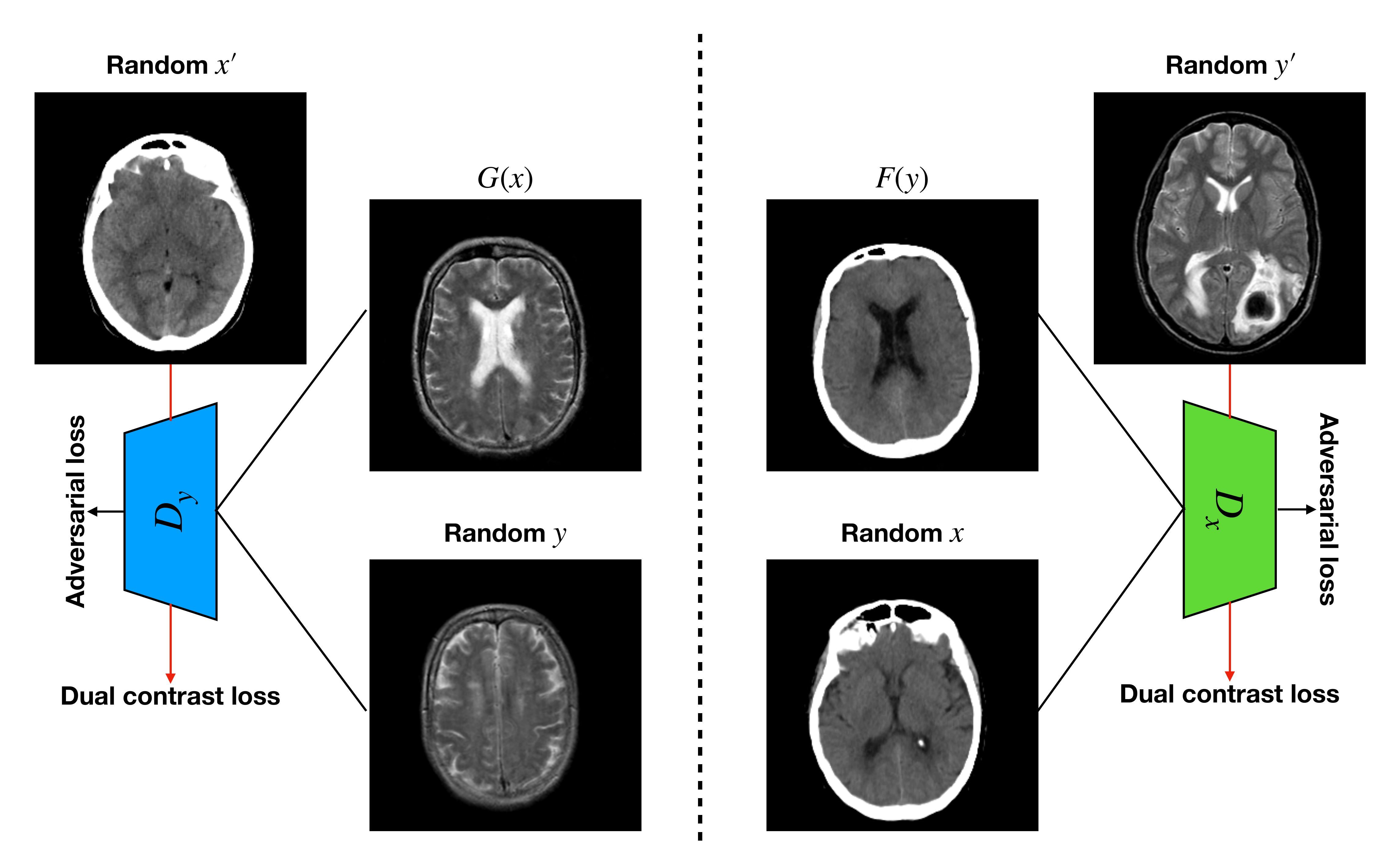}
\caption{The structure of proposed {\color{black}dual contrast}. It adds negative samples, which are randomly selected, from the source domain to avoid mapping current samples close to the negative samples in the latent space.}
\label{trip_image}
\end{figure}

{\color{black}
\subsection{Dual contrast}
\label{Sec:sec:DC}
The original discriminator $D_Y$ aims to distinguish the real images in target domain $y$ from the synthesized images by the generator $G(x)$. 
Although using these images helps the generator to learn features from the $Y$ domain, it can easily fool the discriminator by slightly changing some features of the samples from the source domain
to the target domain, i.e., the discriminator identifies the synthesized image as a real one.  
Thus, the generator $G(x)$ cannot learn a proper mapping in the target domain.   
To alleviate this issue, as shown in Fig.~\ref{trip_image}, we add an additional term, called \textit{dual contrast (DC)}, to leverage the advantage of samples from the source domain $x'$ (see Fig.~\ref{trip_image}), as follows:
\begin{align}
\mathcal{L}_{\mathrm{DC}}\left(D_{Y}, X, Y\right)= \mathbb{E}_{x' \sim p_{\mathrm{data}}(x')}\left[\log (1-D_{Y}\left(x'\right))\right].
\end{align}

In this case, real images are considered class 1, and both synthesized images and samples from the source domain are considered class 0. 
Thus, the discriminator’s goal in our proposed method is to distinguish whether the input image belongs to class 1, i.e., real image, or class 0, i.e., synthesized image or a sample from the source domain. 
Adding images from the source domain as negative samples force the discriminator to guide the generator to synthesize image far away from the source domain in the latent space.
}

\subsection{Training losses}
\label{Sec:sec:Tra}
{\color{black}CycleGAN uses mean squared error (MSE) in its structure to calculate the losses for discriminators, and mean absolute error (MAE) for generators.} However, using MAE or MSE may lead to synthesizing blurry images that cannot obtain human judgments of quality~\cite{snell2017learning}. Besides, the gradient of MSE tends to disappear when it is used in the output layer of neural networks~\cite{zhou2019mpce}, and CE converges fast as its back-propagation error is less than MSE~\cite{sangari2015convergence}. To overcome these issues, in this study, the cross-entropy (CE) and structural similarity index (SSIM)~\cite{Zhou2004Image} are used instead of MSE and MAE, respectively.




\subsubsection{Cross entropy}
\label{Sec:sec:CE}
The main principle behind the CE is to transform the original optimization problem into a stochastic one and then solve it by an adaptive random sampling algorithm~\cite{mannor2005cross}. 
It computes the difference, i.e., distance, between two distributions. CE is a common loss function that has been widely used for classification algorithms. In this study, the CE loss is integrated into the discriminators. Specifically, the PatchGAN's~\cite{isola2017image} discriminator is adopted. It receives an input image, while the output is the probability of being a real image from the target domain. To achieve this, the probability of each patch is calculated, and then the CE loss between the output of the discriminator and the label is computed. Finally, the mean value of all patches is used as the final outcome.
For a binary classification problem, the CE loss with sigmoid activation function can be defined as:   
\begin{equation}\label{CE}
C E(t, y)=-\frac{1}{N} \sum_{i=1}^{N} t_{i} \cdot \log \left(y_{i}\right)+\left(1-t_{i}\right) \cdot \log \left(1-y_{i}\right),
\end{equation}
{\color{black}where $N$ is 256 because of 16$\times$16 patches, $t_{i}$ and $y_{i}$ indicate the label and the predicted probability of the $i$-th patch, respectively.}

\subsubsection{Structural similarity index}
\label{Sec:sec:ssim}
(SSIM)~\cite{Zhou2004Image} is an index that can be used to: \textit{(i)} measure the similarity between two images, e.g., $x_1$ and $x_2$, and \textit{(ii)} evaluate the quality of an image. SSIM, first, normalizes data and then applies a Gaussian filter to each pixel of an image. Therefore, using SSIM can significantly improve the quality of the re-constructed image from another modality in presence of noise as compared with that of MAE~\cite{snell2017learning}.
SSIM considers luminance \emph{l}, contrast \emph{c} and structure \emph{s}, and defined as follows:
\begin{equation}\label{SSIM}
\operatorname{SSIM}(x_{1}, x_{2})=\frac{\left(2 \mu_{x_{1}} \mu_{x_{2}}+c_{1}\right)\left(2 \sigma_{x_{1} x_{2}}+c_{2}\right)}{\left(\mu_{x_{1}}^{2}+\mu_{x_{2}}^{2}+c_{1}\right)\left(\sigma_{x_{1}}^{2}+\sigma_{x_{2}}^{2}+c_{2}\right)},
\end{equation}
where $\mu_{x_{i}}$ and $\sigma_{x_{i}}$ are the mean value and standard deviation of $i$-th image ($i=1,~2$), respectively, $\sigma_{x_{1} x_{2}}$ is the covariance of $x_{1}$ and $x_{2}$. $c_{1}$, $c_{2}$ and $c_{3}$ are constant values, where $c_{3}=\frac{c_{2}}{2}$. In addition, \emph{l}, \emph{c} and \emph{s} can be calculated as follows:
\begin{equation}\label{formulal}
l(x_{1}, x_{2})=\frac{2 \mu_{x_{1}} \mu_{x_{2}}+c_{1}}{\mu_{x_{1}}^{2}+\mu_{x_{2}}^{2}+c_{1}},
\end{equation}
\begin{equation}\label{formulac}
c(x_{1}, x_{2})=\frac{2 \sigma_{x_{1}} \sigma_{x_{2}}+c_{2}}{\sigma_{x_{1}}^{2}+\sigma_{x_{2}}^{2}+c_{2}},
\end{equation}
\begin{equation}\label{formulas}
s(x_{1}, x_{2})=\frac{\sigma_{x_{1} x_{2}}+c_{3}}{\sigma_{x_{1}} \sigma_{x_{2}}+c_{3}}.
\end{equation}

In this study, the SSIM is used for two purposes. First, it is integrated into the generator to synthesis high-quality images from another modality. Second, it is used as an evaluation metric to compute the quality of the generated images.

Finally, by integrating SSIM into the cycle consistency loss, the resulting loss $\mathcal{L}_{\mathrm{scycle}}$ can be written as follows:
\begin{equation}
\label{scycle_consistency_loss}
\mathcal{L}_{\mathrm{scycle}}(G, F) =(1-SSIM(F(G(x)), x))\\
+(1-SSIM(G(F(y)), y)).
\end{equation}

\subsection{Final objective function}
\label{Sec:sec:final}
In this study, the real images from $X$ domain are added as negative samples in $D_Y$ and the real images from $Y$ domain are added as negative samples in $D_X$.  
The final objective function can be written as:
\begin{multline}
\mathcal{L}\left(G, F, D_{X}, D_{Y}\right)= \mathcal{L}_{\mathrm{GAN}}\left(G, D_{Y}, X, Y\right)+\mathcal{L}_{\mathrm{GAN}}\left(F, D_{X}, Y, X\right) \\
+\beta \mathcal{L}_{\mathrm{DC}}\left(D_{Y}, X, Y\right)+\beta \mathcal{L}_{\mathrm{DC}}\left(D_{X}, Y, X\right)
+\lambda \mathcal{L}_{\mathrm{scycle}}(G, F),
\end{multline}
where $\lambda$ and $\beta$ control the weights of cycle consistency and {\color{black}DC losses}, respectively. 
The goal is to optimize:
\begin{align}
G^{*}, F^{*}= \arg \min _{G, F} \max _{D_{X}, D_{Y}} \mathcal{L}\left(G, F, D_{X}, D_{Y}\right).
\end{align}

\section{Experiments}
\label{sec:exp}
This section evaluates the effectiveness of the proposed {\color{black}DC-cycleGAN} model in generating images, i.e., MR from CT or CT from MR, from unpaired data, and compares its performance with other state-of-the-art methods such as cycleGAN~\cite{zhu2017unpaired}, RegGAN~\cite{kong2021breaking}, NICE-GAN~\cite{chen2020reusing}, and DualGAN~\cite{yi2017dualgan}. In addition, an ablation study is conducted to show the effects of using negative samples in synthesizing medical images from other modality. 
In the following subsections, we first present dataset and evaluation metrics and implementation details, then the experimental results along with an ablation study are presented.  


\subsection{Dataset and evaluation metrics}
\label{Sec:sec:daev}

\begin{figure}[tb!]

	\centering
	\subfloat{
		\begin{minipage}[t]{0.24\linewidth}
            \centering
			\includegraphics[width=\textwidth]{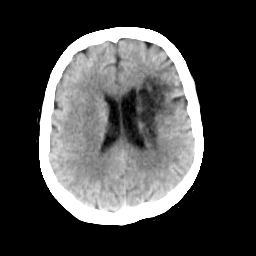}\\
			\vspace{0.1cm}
			\includegraphics[width=\textwidth]{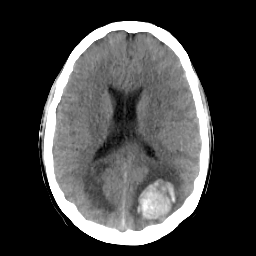}\\
			\vspace{0.1cm}

		\end{minipage}%
	}%
	\subfloat{
		\begin{minipage}[t]{0.24\linewidth}
            \centering
			\includegraphics[width=\textwidth]{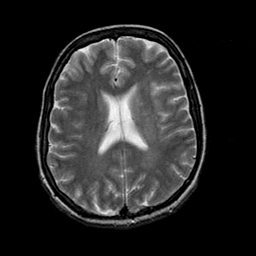}\\
			\vspace{0.1cm}
			\includegraphics[width=\textwidth]{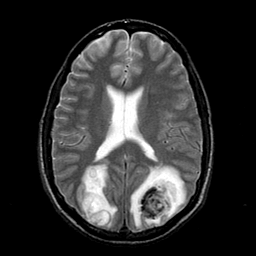}\\
			\vspace{0.1cm}

		\end{minipage}%
	}%
		\subfloat{
		\begin{minipage}[t]{0.24\linewidth}
            \centering
			\includegraphics[width=\textwidth]{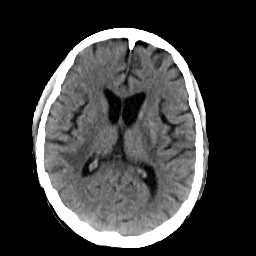}\\
			\vspace{0.1cm}
			\includegraphics[width=\textwidth]{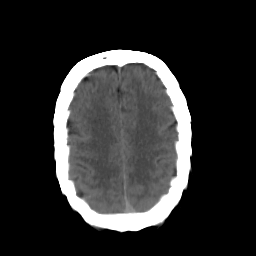}\\
			\vspace{0.1cm}

		\end{minipage}%
	}%
	\subfloat{
		\begin{minipage}[t]{0.24\linewidth}
            \centering
			\includegraphics[width=\textwidth]{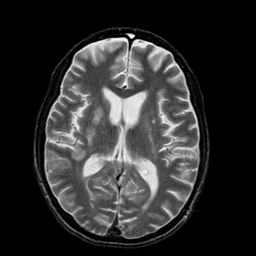}\\
			\vspace{0.1cm}
			\includegraphics[width=\textwidth]{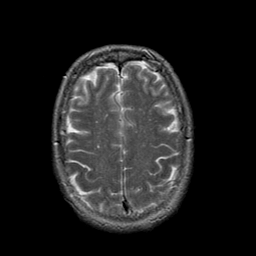}\\
			\vspace{0.1cm}

		\end{minipage}%
	}%

	\centering
	\caption{Four randomly paired data were selected after pre-processing. As can be seen, head frames from CT images and noise from both CT and MR images are removed.  }
	\label{100dataset}
\end{figure}

\subsubsection{Dataset}
\label{Sec:sec:data}
In this study, the dataset introduced in~\cite{han2017mr}, which is publicly available\footnote{https://github.com/ChengBinJin/MRI-to-CT-DCNN-TensorFlow}, is used for performance evaluation. It consists of 367 paired CT and MR images with the size of 512$\times$256. 
In the original dataset, it is noted that several CT images have stereotactic head frame that was used in Gamma Knife treatment. This head frame manually removed from the CT images. In addition, there exist various CT and MR slices that brings incredible difficulty in training the generator. Therefore, 100 images are selected from similar slices for each modality, in which 90 and 10 images are used for training and testing, respectively. Fig.~\ref{100dataset} shows four examples of pre-processed samples.             

\subsubsection{Evaluation metrics}
\label{Sec:sec:ev}
Three performance indicators, including SSIM, mean absolute error (MAE) and peak signal-to-noise ratio (PSNR), are used for performance evaluation and comparison. Among them, SSIM is formulated in subsection~\ref{Sec:sec:ssim}. MAE computes the average absolute between synthetic and real images. It can also be understood as the distance between two pictures. Defined as:  

\begin{equation}
MAE=\frac{1}{M N} \sum_{i=1}^{M} \sum_{j=1}^{N}|x_{1}(i, j)-x_{2}(i, j)|,
\end{equation}
where $M$ and $N$ are high and wide of image, respectively.

PSNR is another performance indicator for image quality assessment that can be computed as follows:
\begin{align}
PSNR=10 \log _{10}(L / \mathrm{MSE}),
\end{align}
where $L$ is dynamic range of the pixel values, and:  
\begin{align}
MSE=\frac{1}{MN} \sum_{i=1}^{M} \sum_{j=1}^{N}(x_{1}(i, j)-x_{2}(i, j))^{2}.
\end{align}

A large PNSR means the real and synthetic images are closer . 

\subsection{Implementation details}
\label{Sec:sec:imp}
Following the same procedure in cycleGAN~\cite{zhu2017unpaired}, the instance normalization~\cite{ulyanov2016instance} is used. The generator contains three convolutions and nine residual blocks for 256 $\times$ 256 and high-resolution images, two fractionally-strided convolutions with stride 0.5.
PatchGANs~\cite{isola2017image} is used as discriminator to identify whether {\color{black}94 x 94} overlapping image patches are real or generated. 
{\color{black}DC-cycleGAN} trains generator five-times and then trains discriminator one-time. 
The parameters of {\color{black}DC-cycleGAN} are set to: $\lambda=10$ and $\beta=0.5$, and the number of epochs and batch sizes are set to 200 and 1, respectively. 
{\color{black}Note that to have a fair comparison with other methods, we followed the original cycleGAN for $\lambda$ and set it to 10. In addition, an experiment is conducted to obtain the best value for $\beta$ (see subsection~\ref{Sec:sec:abl}).} 
These parameters are obtained after several trial-and-errors. The code of {\color{black}DC-cycleGAN} is available online\footnote{https://github.com/JiayuanWang-JW/{\color{black}DC-cycleGAN}}. 
While parameters of other methods are adopted from their references.

{\color{black}For each experiment, we randomly selected 90\% and 10\% of the samples for training and test. This procedure is repeated 10 times and the mean value is used as the final value.}
All images are normalized between -1 and 1, and resized to 256 $\times$ 256. 
All experiments are conducted using a server with Intel(R) Xeon(R) E5-2650 CPU and Nvidia GTX 1080TI GPU.

\subsection{Results comparison}
\label{Sec:sec:res}
In this section, a comprehensive comparison between our proposed {\color{black}DC-cycleGAN} with other recently published methods is conducted. These methods include conventional cycleGAN~\cite{zhu2017unpaired}, RegGAN~\cite{kong2021breaking}, NICE-GAN~\cite{chen2020reusing} and DualGAN~\cite{yi2017dualgan}. Their published codes\footnote{https://github.com/simontomaskarlsson/CycleGAN-Keras} \footnote{https://github.com/kid-liet/reg-gan} \footnote{https://github.com/alpc91/NICE-GAN-pytorch} \footnote{https://github.com/duxingren14/DualGAN} are used to generate results using the same parameters in their corresponding references.  

\begin{table}[hbt!]
\centering
\caption{MR synthesis quality evaluation metrics ``Mean (standard deviation)" for various methods.}
\begin{adjustbox} {width=\columnwidth}
\label{CT-MRI result}
\begin{tabular}{l c c c c}
  \toprule
  \multirow{1}*{Method} & \multicolumn{1}{c}{MAE} & \multicolumn{1}{c}{PSNR} & \multicolumn{1}{c}{SSIM}\\
  
  \midrule
        RegGAN~\cite{kong2021breaking} & 0.16828 (0.14436)  & 18.66927 (7.14952)  & 0.68031 (0.14813)   \\ 
        CycleGAN~\cite{zhu2017unpaired} & 0.09155 (0.02325)  & 20.63825 (2.28951)  & 0.71670 (0.04504)   \\
        NICE-GAN~\cite{chen2020reusing} & 0.08148 (0.00184) & 21.44107 (0.21873) & 0.70288 (0.00579)    \\
        DualGAN~\cite{yi2017dualgan} & 0.08038 (0.03109) & 22.84981 (3.83697) &	0.74028 (0.08261)\\
        {\color{black}DC-cycleGAN} & \textbf{0.04559} (0.00333)  & \textbf{26.68858} (0.82837)  & \textbf{0.82622} (0.01220)   \\ 
  \bottomrule
\end{tabular}
\end{adjustbox}
\end{table}

\begin{table}[hbt!]
\centering
\caption{CT synthesis quality evaluation metrics ``Mean (standard deviation)" for various methods.}
\begin{adjustbox} {width=\columnwidth}
\label{MRI-CT result}
\begin{tabular}{l c c c c}
  \toprule
  \multirow{1}*{Method} & \multicolumn{1}{c}{MAE} & \multicolumn{1}{c}{PSNR} & \multicolumn{1}{c}{SSIM}\\

  \midrule
        RegGAN~\cite{kong2021breaking} & 0.04741 (0.00406)  & 19.88008 (0.63692)  & 0.81250 (0.00678)   \\ 
        CycleGAN~\cite{zhu2017unpaired} & 0.05974 (0.01193)  & 20.07402 (1.52217)  & 0.81600 (0.02008)   \\
        NICE-GAN~\cite{chen2020reusing} & 0.08373 (0.00789) & 17.03886 (0.56989) & 0.78790 (0.01014)    \\
        DualGAN~\cite{yi2017dualgan} & 0.04890 (0.01515) & 22.11585 (1.60191) &	0.83658 (0.02182)  \\ 
        {\color{black}DC-cycleGAN} & \textbf{0.03483} (0.00377)  & \textbf{25.19949} (1.67149)  & \textbf{0.87035} (0.01210)   \\
  \bottomrule
\end{tabular}
\end{adjustbox}
\end{table}


Tables~\ref{CT-MRI result} and~\ref{MRI-CT result} show the quantitative accuracies of various methods in generating CT and MR images, respectively. As can be seen, {\color{black}DC-cycleGAN} outperforms other methods in terms of MAE, PNSR and SSIM. This is mainly due to the {\color{black}dual contrast} loss that is integrated into the {\color{black}DC-cycleGAN} structure. 

\begin{figure}[htb!]
	\centering
	\subfloat[]{
		\begin{minipage}[t]{0.16\textwidth}
			\centering
			\includegraphics[width=0.95\textwidth]{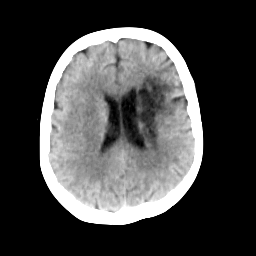}\\
			\vspace{0.1cm}
			\includegraphics[width=0.95\textwidth]{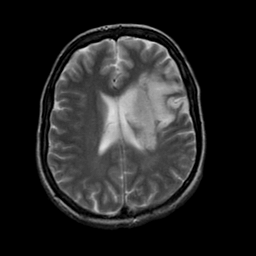}\\
			\vspace{0.4cm}
		\end{minipage}%
	}%
	\subfloat[]{
		\begin{minipage}[t]{0.16\textwidth}
			\centering
			\includegraphics[width=0.95\textwidth]{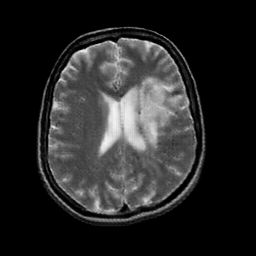}\\
			\vspace{0.1cm}
			\includegraphics[width=\textwidth]{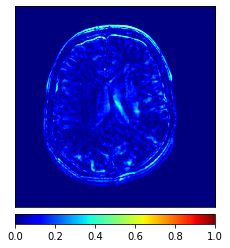}\\
			\vspace{0.1cm}
		\end{minipage}%
	}%
	\subfloat[]{
		\begin{minipage}[t]{0.16\textwidth}
			\centering
			\includegraphics[width=0.95\textwidth]{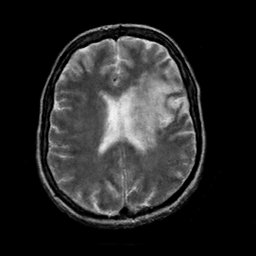}\\
			\vspace{0.1cm}
			\includegraphics[width=\textwidth]{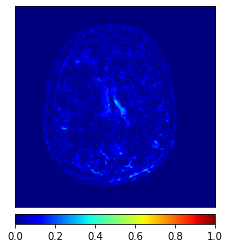}\\
			\vspace{0.1cm}
		\end{minipage}%
	}%
	\subfloat[]{
		\begin{minipage}[t]{0.16\textwidth}
			\centering
			\includegraphics[width=0.95\textwidth]{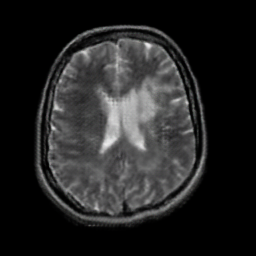}\\
			\vspace{0.1cm}
			\includegraphics[width=\textwidth]{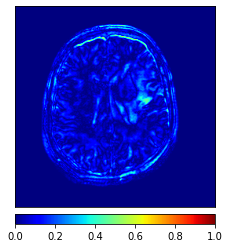}\\
			\vspace{0.1cm}
		\end{minipage}%
	}%
	\subfloat[]{
		\begin{minipage}[t]{0.16\textwidth}
			\centering
			\includegraphics[width=0.95\textwidth]{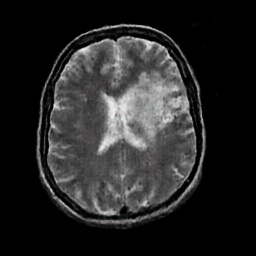}\\
			\vspace{0.1cm}
			\includegraphics[width=\textwidth]{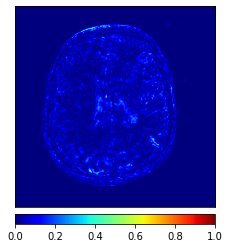}\\
			\vspace{0.1cm}
		\end{minipage}%
	}%
	\subfloat[]{
		\begin{minipage}[t]{0.16\textwidth}
			\centering
			\includegraphics[width=0.95\textwidth]{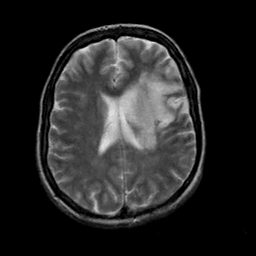}\\
			\vspace{0.1cm}
			\includegraphics[width=\textwidth]{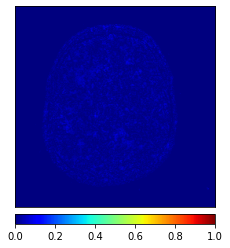}\\
			\vspace{0.1cm}
		\end{minipage}%
	}%

	\centering
	\caption{Synthesized MR images along with the absolute error maps between groundtruth and synthesized images by different methods. (a) Real image, (b) RegGAN, (c) CycleGAN, (d) NICE-GAN, (e) DualGAN, (f) DC-cycleGAN}
	\label{comparison result}
\end{figure}

\begin{figure}[htb!]
	\centering
	\subfloat[]{
		\begin{minipage}[t]{0.16\textwidth}
			\centering
			\includegraphics[width=0.95\textwidth]{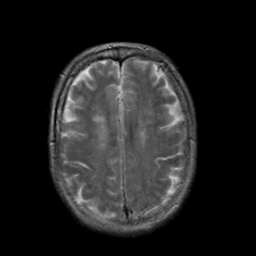}\\
			\vspace{0.1cm}
			\includegraphics[width=0.95\textwidth]{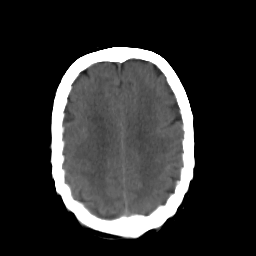}\\
			\vspace{0.4cm}
		\end{minipage}%
	}%
	\subfloat[]{
		\begin{minipage}[t]{0.16\textwidth}
			\centering
			\includegraphics[width=0.95\textwidth]{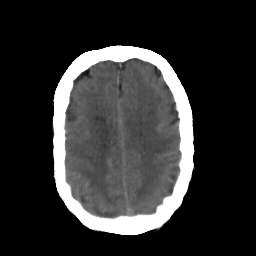}\\
			\vspace{0.1cm}
			\includegraphics[width=\textwidth]{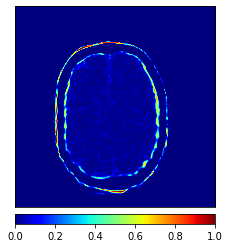}\\
			\vspace{0.1cm}
		\end{minipage}%
	}%
	\subfloat[]{
		\begin{minipage}[t]{0.16\textwidth}
			\centering
			\includegraphics[width=0.95\textwidth]{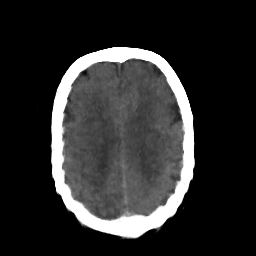}\\
			\vspace{0.1cm}
			\includegraphics[width=\textwidth]{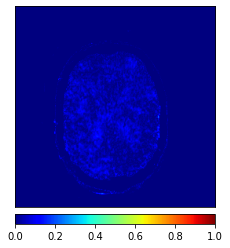}\\
			\vspace{0.1cm}
		\end{minipage}%
	}%
	\subfloat[]{
		\begin{minipage}[t]{0.16\textwidth}
			\centering
			\includegraphics[width=0.95\textwidth]{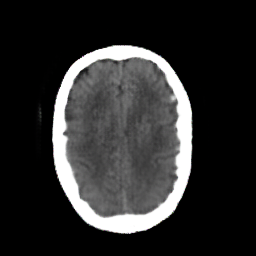}\\
			\vspace{0.1cm}
			\includegraphics[width=\textwidth]{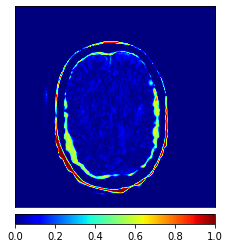}\\
			\vspace{0.1cm}
		\end{minipage}%
	}%
	\subfloat[]{
		\begin{minipage}[t]{0.16\textwidth}
			\centering
			\includegraphics[width=0.95\textwidth]{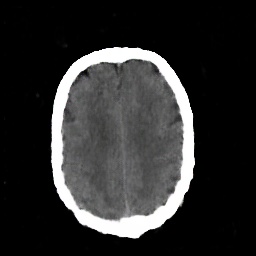}\\
			\vspace{0.1cm}
			\includegraphics[width=\textwidth]{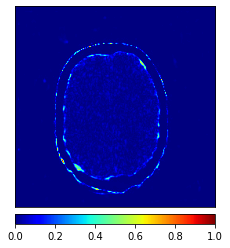}\\
			\vspace{0.1cm}
		\end{minipage}%
	}%
	\subfloat[]{
		\begin{minipage}[t]{0.16\textwidth}
			\centering
			\includegraphics[width=0.95\textwidth]{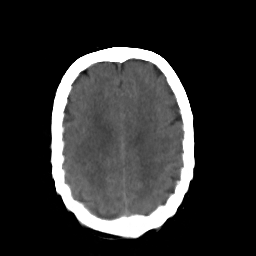}\\
			\vspace{0.1cm}
			\includegraphics[width=\textwidth]{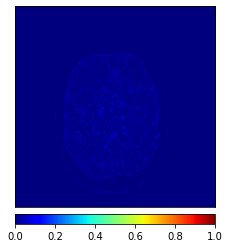}\\
			\vspace{0.1cm}
		\end{minipage}%
	}%

	\centering
	\caption{Synthesized CT images along with the absolute error maps between groundtruth and synthesized images by different methods. (a) Real image, (b) RegGAN, (c) CycleGAN, (d) NICE-GAN, (e)DualGAN, (f) DC-cycleGAN}
	\label{visualCT}
\end{figure}

Moreover, Figs.~\ref{comparison result} and~\ref{visualCT} show the synthesized MR and CT images along with the errors between the real and synthesized images by different methods, respectively. It can be seen that the synthesized images by {\color{black}DC-cycleGAN} are more identical to the real ones as compared with other methods. This indicates that effectiveness of SSIM and CE along with {\color{black}dual contrast} in synthesizing images. In addition, the error between the groundtruth and synthesized MR/CT images by {\color{black}DC-cycleGAN} is relatively less as compared with other methods.    

\subsection{Ablation study and sensitivity analysis}
\label{Sec:sec:abl}
This section conducts an ablation study to show the impacts of using {\color{black}dual contrast}, SSIM and CE in the {\color{black}DC-cycleGAN} structure. To achieve this, four scenarios are considered, including:
\begin{itemize}
    \item MAE and MSE without {\color{black}dual contrast} denoted as MAE \& MSE (wo),
    \item MAE and MSE with {\color{black}dual contrast} denoted as MAE \& MSE (w),
    \item SSIM and CE without {\color{black}dual contrast} denoted as SSIM \& CE (wo), and
    \item SSIM and CE with {\color{black}dual contrast} denoted as SSIM \& CE (w).
\end{itemize}

\begin{figure*}
	\centering
	\subfloat{
		\begin{minipage}[t]{0.33\textwidth}
			\centering
			\includegraphics[width=\textwidth]{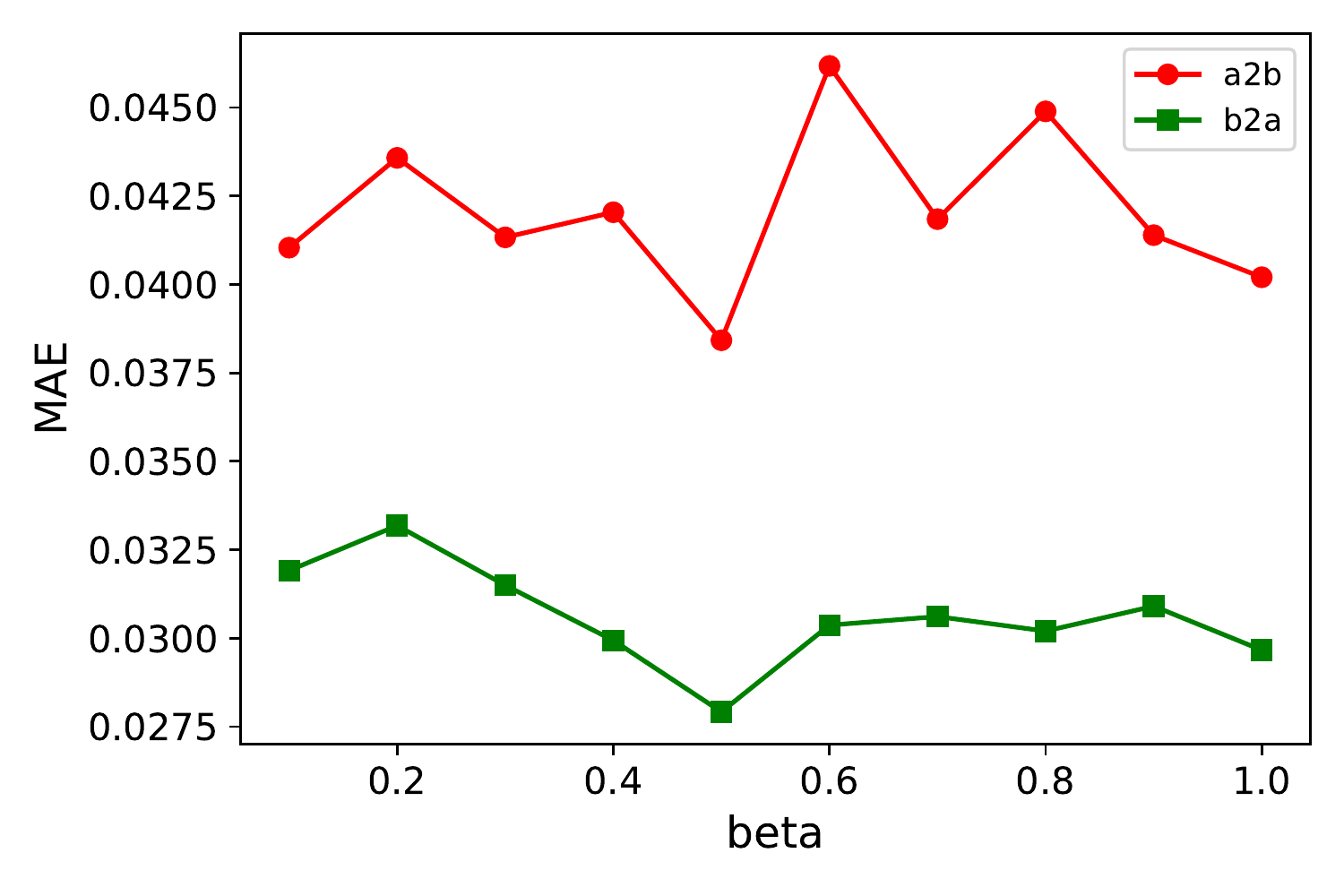}\\
		\end{minipage}%
		\begin{minipage}[t]{0.33\textwidth}
			\centering
			\includegraphics[width=\textwidth]{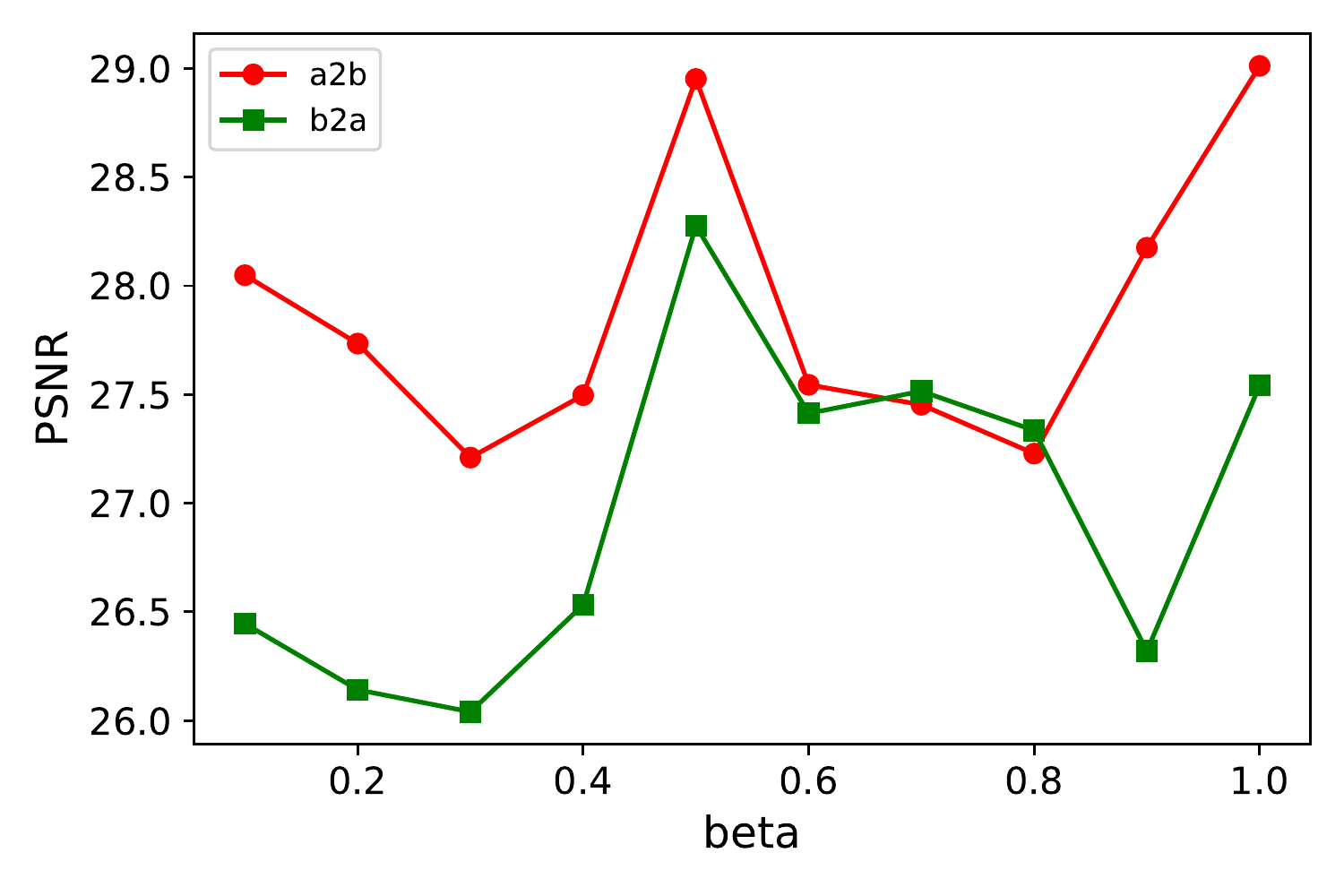}\\
		\end{minipage}%
		\begin{minipage}[t]{0.33\textwidth}
			\centering
			\includegraphics[width=\textwidth]{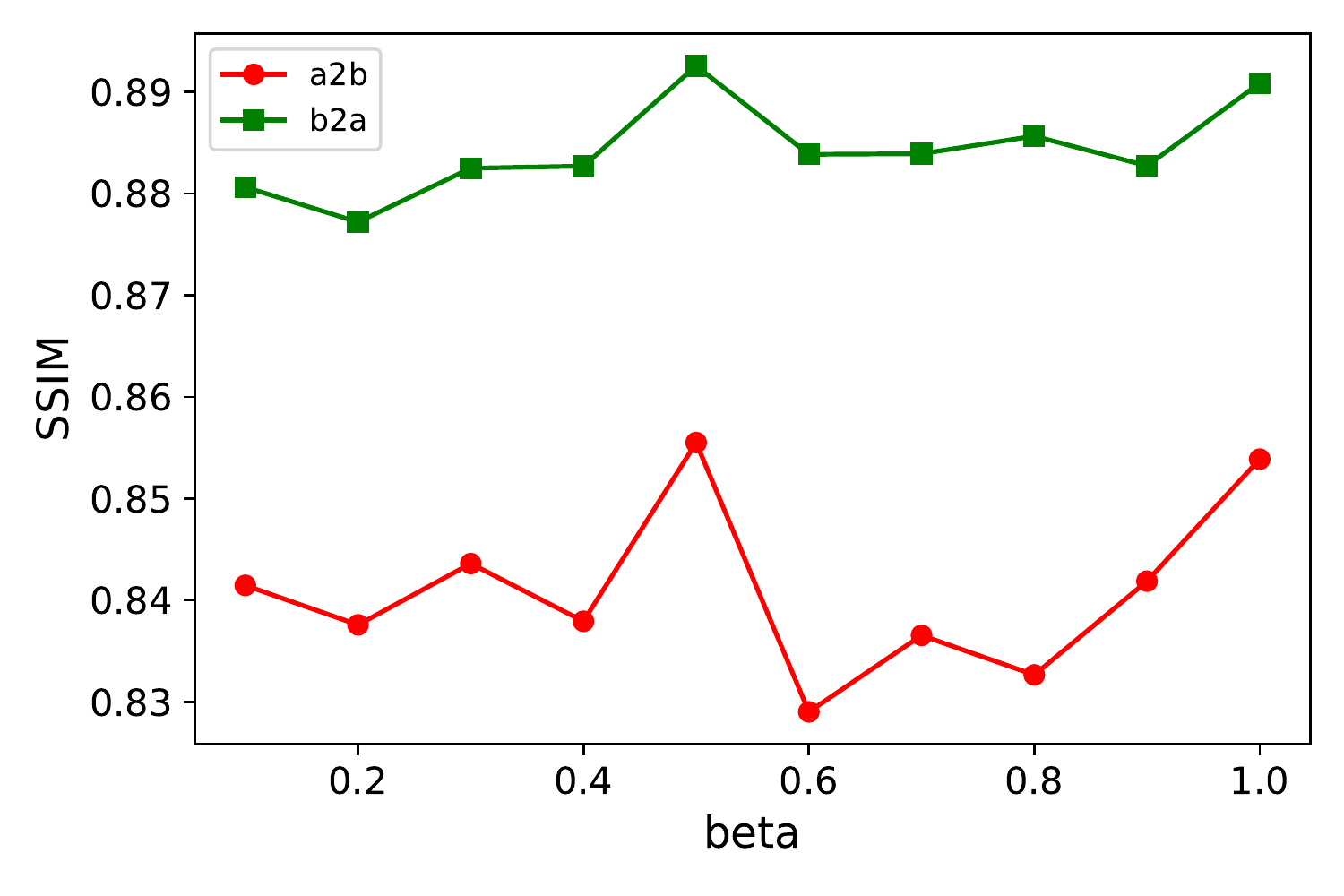}\\
		\end{minipage}%
	}%
	\centering
	\caption{The sensitivity analysis for $\beta$. ``a2b" and ``b2a" indicate CT-to-MR and MR-to-CT, respectively. }
	\label{sensitive}
\end{figure*}

{\color{black}Tables~\ref{ablation_CT-MRI} and~\ref{ablation_MRI-CT} show the results of MR and CT synthesis, respectively. As can be seen, all components play vital role in both tables. 
SSIM \& CE (w) performs significantly better than other losses in synthesizing MR images. This also can be seen visually in Fig.~\ref{MAEMSE_cessim_a2b}. 
In contrast, SSIM\&CE (w) performs slightly better than SSIM\&CE (wo) in synthesizing CT images, both generate more or less similar CT images (see Table~\ref{ablation_MRI-CT}).\par
\begin{table}[hbt!]
\centering
\caption{ Ablation study. MR synthesis quality evaluation metrics ``Mean (standard deviation)" under different conditions. ``(w)" and ``(wo)" indicate with and without {\color{black}dual contrast}, respectively.}
\begin{adjustbox} {width=\columnwidth}
\label{ablation_CT-MRI}
\begin{tabular}{l c c c c}
  \toprule
  \multirow{1}*{Method} & \multicolumn{1}{c}{MAE} & \multicolumn{1}{c}{PSNR} & \multicolumn{1}{c}{SSIM}\\
  \midrule
        MAE \& MSE (wo) & 0.06850 (0.01598) & 23.62560 (2.42909) & 0.78583 (0.03267)    \\
        MAE \& MSE (w) & 0.06939 (0.01849)  & 23.79743 (2.92001)  & 0.78368 (0.03924)   \\ 
        SSIM \& CE (wo) & 0.05562 (0.01858)  & 25.53608 (2.01767)  & 0.80613 (0.03585)   \\
        SSIM \& CE (w) & \textbf{0.04559} (0.00333)  & \textbf{26.68858} (0.82837)  & \textbf{0.82622} (0.01220)   \\ 
  \bottomrule
\end{tabular}
\end{adjustbox}
\end{table}

\begin{table}[hbt!]
\centering
\caption{Ablation study. CT synthesis quality evaluation metrics ``Mean (standard deviation)" under different conditions. ``(w)" and ``(wo)" indicate with and without {\color{black}dual contrast}, respectively.}
\begin{adjustbox} {width=\columnwidth}
\label{ablation_MRI-CT}
\begin{tabular}{l c c c c}
  \toprule
  \multirow{1}*{Method} & \multicolumn{1}{c}{MAE} & \multicolumn{1}{c}{PSNR} & \multicolumn{1}{c}{SSIM}\\
  \midrule
        MAE\&MSE (wo) & 0.04172 (0.01401)  & 23.57607 (2.93920)  & 0.85799 (0.02542)   \\
        MAE\&MSE (w) & 0.03800 (0.00721)  & 24.24584 (2.34511)  & 0.86124 (0.01885)   \\ 
        SSIM\&CE (wo) & 0.03582 (0.00400)  & 24.42528 (2.12593)  & 0.86676 (0.01117)   \\
        SSIM\&CE (w) & \textbf{0.03483} (0.00377)  & \textbf{25.19949} (1.67149)  & \textbf{0.87035} (0.01210)   \\ 
  \bottomrule
\end{tabular}
\end{adjustbox}
\end{table}

In addition, we have conducted an experiment to show the sensitivity of DC-cycleGAN to $\beta$. To achieve this, $\beta$ is varied from 0.1 to 1. As shown in Fig.~\ref{sensitive}, the DC-cycleGAN produces the best results for both directions when $\beta$ is set to 0.5.

}



\section{Conclusion}
\label{Sec:con}
In this study, the {\color{black}DC-cycleGAN} model has been proposed for unsupervised medical image synthesis. It leverages the advantage of samples from the source domain as negative samples in learning a mapping between two image modalities using a {\color{black}dual contrast} loss. It maps the learning samples far away from the images of the source domain. In addition, SSIM and CE losses are integrated into the {\color{black}DC-cycleGAN} structure to consider both luminance and structure of samples.      
{\color{black}DC-cycleGAN} is evaluated on brain MR-to-CT images and the results indicate that {\color{black}DC-cycleGAN} is able to produce better CT/MR images in terms of both accuracy and visual quality as compared with state-of-the-art methods such as cycleGAN, RegGAN and dualGAN. Moreover, an ablation study is conducted to evaluate the impacts of different parts on the performance.
One limitation of this study is using images from one slice to train {\color{black}DC-cycleGAN} and other methods, which is not realistic in practice. Our future work is focused on developing medical image synthesis models that are able to learn from datasets with multiple slices.





\bibliographystyle{cas-model2-names}
\bibliography{ref.bib}

\begin{thebibliography}{49}
\expandafter\ifx\csname natexlab\endcsname\relax\def\natexlab#1{#1}\fi
\providecommand{\url}[1]{\texttt{#1}}
\providecommand{\href}[2]{#2}
\providecommand{\path}[1]{#1}
\providecommand{\DOIprefix}{doi:}
\providecommand{\ArXivprefix}{arXiv:}
\providecommand{\URLprefix}{URL: }
\providecommand{\Pubmedprefix}{pmid:}
\providecommand{\doi}[1]{\href{http://dx.doi.org/#1}{\path{#1}}}
\providecommand{\Pubmed}[1]{\href{pmid:#1}{\path{#1}}}
\providecommand{\bibinfo}[2]{#2}
\ifx\xfnm\relax \def\xfnm[#1]{\unskip,\space#1}\fi
\bibitem[{Abu-Srhan et~al.(2021)Abu-Srhan, Almallahi, Abushariah, Mahafza and
  Al-Kadi}]{abu2021paired}
\bibinfo{author}{Abu-Srhan, A.}, \bibinfo{author}{Almallahi, I.},
  \bibinfo{author}{Abushariah, M.A.}, \bibinfo{author}{Mahafza, W.},
  \bibinfo{author}{Al-Kadi, O.S.}, \bibinfo{year}{2021}.
\newblock \bibinfo{title}{Paired-unpaired unsupervised attention guided gan
  with transfer learning for bidirectional brain mr-ct synthesis}.
\newblock \bibinfo{journal}{Computers in Biology and Medicine}
  \bibinfo{volume}{136}, \bibinfo{pages}{104763}.
\bibitem[{Bi et~al.(2017)Bi, Kim, Kumar, Feng and Fulham}]{bi2017synthesis}
\bibinfo{author}{Bi, L.}, \bibinfo{author}{Kim, J.}, \bibinfo{author}{Kumar,
  A.}, \bibinfo{author}{Feng, D.}, \bibinfo{author}{Fulham, M.},
  \bibinfo{year}{2017}.
\newblock \bibinfo{title}{Synthesis of positron emission tomography (pet)
  images via multi-channel generative adversarial networks (gans)}, in:
  \bibinfo{booktitle}{molecular imaging, reconstruction and analysis of moving
  body organs, and stroke imaging and treatment}, pp. \bibinfo{pages}{43--51}.
\bibitem[{Chartsias et~al.(2017)Chartsias, Joyce, Dharmakumar and
  Tsaftaris}]{chartsias2017adversarial}
\bibinfo{author}{Chartsias, A.}, \bibinfo{author}{Joyce, T.},
  \bibinfo{author}{Dharmakumar, R.}, \bibinfo{author}{Tsaftaris, S.A.},
  \bibinfo{year}{2017}.
\newblock \bibinfo{title}{Adversarial image synthesis for unpaired multi-modal
  cardiac data}, in: \bibinfo{booktitle}{International workshop on simulation
  and synthesis in medical imaging}, \bibinfo{organization}{Springer}. pp.
  \bibinfo{pages}{3--13}.
\bibitem[{Chen et~al.(2020)Chen, Huang, Huang, Sun and Fang}]{chen2020reusing}
\bibinfo{author}{Chen, R.}, \bibinfo{author}{Huang, W.},
  \bibinfo{author}{Huang, B.}, \bibinfo{author}{Sun, F.},
  \bibinfo{author}{Fang, B.}, \bibinfo{year}{2020}.
\newblock \bibinfo{title}{Reusing discriminators for encoding: Towards
  unsupervised image-to-image translation}, in: \bibinfo{booktitle}{Proceedings
  of the IEEE/CVF Conference on Computer Vision and Pattern Recognition}, pp.
  \bibinfo{pages}{8168--8177}.
\bibitem[{Chen et~al.(2018)Chen, Qin, Zhou and Yan}]{chen2018u}
\bibinfo{author}{Chen, S.}, \bibinfo{author}{Qin, A.}, \bibinfo{author}{Zhou,
  D.}, \bibinfo{author}{Yan, D.}, \bibinfo{year}{2018}.
\newblock \bibinfo{title}{U-net-generated synthetic ct images for magnetic
  resonance imaging-only prostate intensity-modulated radiation therapy
  treatment planning}.
\newblock \bibinfo{journal}{Medical physics} \bibinfo{volume}{45},
  \bibinfo{pages}{5659--5665}.
\bibitem[{Fu et~al.(2018)Fu, Yang, Singhrao, Ruan, Low and Lewis}]{fu2018male}
\bibinfo{author}{Fu, J.}, \bibinfo{author}{Yang, Y.},
  \bibinfo{author}{Singhrao, K.}, \bibinfo{author}{Ruan, D.},
  \bibinfo{author}{Low, D.A.}, \bibinfo{author}{Lewis, J.H.},
  \bibinfo{year}{2018}.
\newblock \bibinfo{title}{Male pelvic synthetic ct generation from t1-weighted
  mri using 2d and 3d convolutional neural networks}.
\newblock \bibinfo{journal}{arXiv preprint arXiv:1803.00131} .
\bibitem[{Goodfellow et~al.(2014)Goodfellow, Pouget-Abadie, Mirza, Xu,
  Warde-Farley, Ozair, Courville and Bengio}]{goodfellow2014generative}
\bibinfo{author}{Goodfellow, I.}, \bibinfo{author}{Pouget-Abadie, J.},
  \bibinfo{author}{Mirza, M.}, \bibinfo{author}{Xu, B.},
  \bibinfo{author}{Warde-Farley, D.}, \bibinfo{author}{Ozair, S.},
  \bibinfo{author}{Courville, A.}, \bibinfo{author}{Bengio, Y.},
  \bibinfo{year}{2014}.
\newblock \bibinfo{title}{Generative adversarial nets}, in:
  \bibinfo{booktitle}{Advances in Neural Information Processing Systems}, pp.
  \bibinfo{pages}{2672--2680}.
\bibitem[{Han(2017)}]{han2017mr}
\bibinfo{author}{Han, X.}, \bibinfo{year}{2017}.
\newblock \bibinfo{title}{Mr-based synthetic ct generation using a deep
  convolutional neural network method}.
\newblock \bibinfo{journal}{Medical physics} \bibinfo{volume}{44},
  \bibinfo{pages}{1408--1419}.
\bibitem[{Heinrich et~al.(2012)Heinrich, Jenkinson, Bhushan, Matin, Gleeson,
  Brady and Schnabel}]{heinrich2012mind}
\bibinfo{author}{Heinrich, M.P.}, \bibinfo{author}{Jenkinson, M.},
  \bibinfo{author}{Bhushan, M.}, \bibinfo{author}{Matin, T.},
  \bibinfo{author}{Gleeson, F.V.}, \bibinfo{author}{Brady, M.},
  \bibinfo{author}{Schnabel, J.A.}, \bibinfo{year}{2012}.
\newblock \bibinfo{title}{Mind: Modality independent neighbourhood descriptor
  for multi-modal deformable registration}.
\newblock \bibinfo{journal}{Medical image analysis} \bibinfo{volume}{16},
  \bibinfo{pages}{1423--1435}.
\bibitem[{Hiasa et~al.(2018)Hiasa, Otake, Takao, Matsuoka, Takashima, Carass,
  Prince, Sugano and Sato}]{hiasa2018cross}
\bibinfo{author}{Hiasa, Y.}, \bibinfo{author}{Otake, Y.},
  \bibinfo{author}{Takao, M.}, \bibinfo{author}{Matsuoka, T.},
  \bibinfo{author}{Takashima, K.}, \bibinfo{author}{Carass, A.},
  \bibinfo{author}{Prince, J.L.}, \bibinfo{author}{Sugano, N.},
  \bibinfo{author}{Sato, Y.}, \bibinfo{year}{2018}.
\newblock \bibinfo{title}{Cross-modality image synthesis from unpaired data
  using cyclegan}, in: \bibinfo{booktitle}{International workshop on simulation
  and synthesis in medical imaging}, \bibinfo{organization}{Springer}. pp.
  \bibinfo{pages}{31--41}.
\bibitem[{Hsu et~al.(2013)Hsu, Cao, Huang, Feng and
  Balter}]{hsu2013investigation}
\bibinfo{author}{Hsu, S.H.}, \bibinfo{author}{Cao, Y.}, \bibinfo{author}{Huang,
  K.}, \bibinfo{author}{Feng, M.}, \bibinfo{author}{Balter, J.M.},
  \bibinfo{year}{2013}.
\newblock \bibinfo{title}{Investigation of a method for generating synthetic ct
  models from mri scans of the head and neck for radiation therapy}.
\newblock \bibinfo{journal}{Physics in Medicine \& Biology}
  \bibinfo{volume}{58}, \bibinfo{pages}{8419}.
\bibitem[{Isola et~al.(2017)Isola, Zhu, Zhou and Efros}]{isola2017image}
\bibinfo{author}{Isola, P.}, \bibinfo{author}{Zhu, J.Y.},
  \bibinfo{author}{Zhou, T.}, \bibinfo{author}{Efros, A.A.},
  \bibinfo{year}{2017}.
\newblock \bibinfo{title}{Image-to-image translation with conditional
  adversarial networks}, in: \bibinfo{booktitle}{Proceedings of the IEEE
  conference on computer vision and pattern recognition}, pp.
  \bibinfo{pages}{1125--1134}.
\bibitem[{Kang et~al.(2021)Kang, An, Jin, Kim, Chie, Park and
  Lee}]{SeungKwan2021Synthetic}
\bibinfo{author}{Kang, S.K.}, \bibinfo{author}{An, H.J.}, \bibinfo{author}{Jin,
  H.}, \bibinfo{author}{Kim, J.i.}, \bibinfo{author}{Chie, E.K.},
  \bibinfo{author}{Park, J.M.}, \bibinfo{author}{Lee, J.S.},
  \bibinfo{year}{2021}.
\newblock \bibinfo{title}{Synthetic ct generation from weakly paired mr images
  using cycle-consistent gan for mr-guided radiotherapy}.
\newblock \bibinfo{journal}{Biomedical Engineering Letters} ,
  \bibinfo{pages}{1--9}.
\bibitem[{Kingma and Welling(2013)}]{kingma2013auto}
\bibinfo{author}{Kingma, D.P.}, \bibinfo{author}{Welling, M.},
  \bibinfo{year}{2013}.
\newblock \bibinfo{title}{Auto-encoding variational bayes}.
\newblock \bibinfo{journal}{arXiv preprint arXiv:1312.6114} .
\bibitem[{Kong et~al.(2021)Kong, Lian, Huang, Hu, Zhou
  et~al.}]{kong2021breaking}
\bibinfo{author}{Kong, L.}, \bibinfo{author}{Lian, C.}, \bibinfo{author}{Huang,
  D.}, \bibinfo{author}{Hu, Y.}, \bibinfo{author}{Zhou, Q.}, et~al.,
  \bibinfo{year}{2021}.
\newblock \bibinfo{title}{Breaking the dilemma of medical image-to-image
  translation}.
\newblock \bibinfo{journal}{Advances in Neural Information Processing Systems}
  \bibinfo{volume}{34}.
\bibitem[{Krizhevsky et~al.(2012)Krizhevsky, Sutskever and
  Hinton}]{krizhevsky2012imagenet}
\bibinfo{author}{Krizhevsky, A.}, \bibinfo{author}{Sutskever, I.},
  \bibinfo{author}{Hinton, G.E.}, \bibinfo{year}{2012}.
\newblock \bibinfo{title}{Imagenet classification with deep convolutional
  neural networks}.
\newblock \bibinfo{journal}{Advances in neural information processing systems}
  \bibinfo{volume}{25}.
\bibitem[{Mannor et~al.(2005)Mannor, Peleg and Rubinstein}]{mannor2005cross}
\bibinfo{author}{Mannor, S.}, \bibinfo{author}{Peleg, D.},
  \bibinfo{author}{Rubinstein, R.}, \bibinfo{year}{2005}.
\newblock \bibinfo{title}{The cross entropy method for classification}, in:
  \bibinfo{booktitle}{Proceedings of the 22nd international conference on
  Machine learning}, pp. \bibinfo{pages}{561--568}.
\bibitem[{Mastrogiacomo et~al.(2019)Mastrogiacomo, Dou, Jansen and
  Walboomers}]{mastrogiacomo2019magnetic}
\bibinfo{author}{Mastrogiacomo, S.}, \bibinfo{author}{Dou, W.},
  \bibinfo{author}{Jansen, J.A.}, \bibinfo{author}{Walboomers, X.F.},
  \bibinfo{year}{2019}.
\newblock \bibinfo{title}{Magnetic resonance imaging of hard tissues and hard
  tissue engineered bio-substitutes}.
\newblock \bibinfo{journal}{Molecular imaging and biology}
  \bibinfo{volume}{21}, \bibinfo{pages}{1003--1019}.
\bibitem[{M{\'e}rida et~al.(2015)M{\'e}rida, Costes, Heckemann, Drzezga,
  F{\"o}rster and Hammers}]{merida2015evaluation}
\bibinfo{author}{M{\'e}rida, I.}, \bibinfo{author}{Costes, N.},
  \bibinfo{author}{Heckemann, R.A.}, \bibinfo{author}{Drzezga, A.},
  \bibinfo{author}{F{\"o}rster, S.}, \bibinfo{author}{Hammers, A.},
  \bibinfo{year}{2015}.
\newblock \bibinfo{title}{Evaluation of several multi-atlas methods for
  pseudo-ct generation in brain mri-pet attenuation correction}, in:
  \bibinfo{booktitle}{2015 IEEE 12th international symposium on biomedical
  imaging (ISBI)}, \bibinfo{organization}{IEEE}. pp.
  \bibinfo{pages}{1431--1434}.
\bibitem[{Nie et~al.(2016)Nie, Cao, Gao, Wang and Shen}]{nie2016estimating}
\bibinfo{author}{Nie, D.}, \bibinfo{author}{Cao, X.}, \bibinfo{author}{Gao,
  Y.}, \bibinfo{author}{Wang, L.}, \bibinfo{author}{Shen, D.},
  \bibinfo{year}{2016}.
\newblock \bibinfo{title}{Estimating ct image from mri data using 3d fully
  convolutional networks}, in: \bibinfo{booktitle}{Deep Learning and Data
  Labeling for Medical Applications}, pp. \bibinfo{pages}{170--178}.
\bibitem[{Nie et~al.(2017)Nie, Trullo, Lian, Petitjean, Ruan, Wang and
  Shen}]{nie2017medical}
\bibinfo{author}{Nie, D.}, \bibinfo{author}{Trullo, R.}, \bibinfo{author}{Lian,
  J.}, \bibinfo{author}{Petitjean, C.}, \bibinfo{author}{Ruan, S.},
  \bibinfo{author}{Wang, Q.}, \bibinfo{author}{Shen, D.}, \bibinfo{year}{2017}.
\newblock \bibinfo{title}{Medical image synthesis with context-aware generative
  adversarial networks}, in: \bibinfo{booktitle}{International conference on
  medical image computing and computer-assisted intervention},
  \bibinfo{organization}{Springer}. pp. \bibinfo{pages}{417--425}.
\bibitem[{Nie et~al.(2018)Nie, Trullo, Lian, Wang, Petitjean, Ruan, Wang and
  Shen}]{nie2018medical}
\bibinfo{author}{Nie, D.}, \bibinfo{author}{Trullo, R.}, \bibinfo{author}{Lian,
  J.}, \bibinfo{author}{Wang, L.}, \bibinfo{author}{Petitjean, C.},
  \bibinfo{author}{Ruan, S.}, \bibinfo{author}{Wang, Q.},
  \bibinfo{author}{Shen, D.}, \bibinfo{year}{2018}.
\newblock \bibinfo{title}{Medical image synthesis with deep convolutional
  adversarial networks}.
\newblock \bibinfo{journal}{IEEE Transactions on Biomedical Engineering}
  \bibinfo{volume}{65}, \bibinfo{pages}{2720--2730}.
\bibitem[{Odena et~al.(2017)Odena, Olah and Shlens}]{odena2017conditional}
\bibinfo{author}{Odena, A.}, \bibinfo{author}{Olah, C.},
  \bibinfo{author}{Shlens, J.}, \bibinfo{year}{2017}.
\newblock \bibinfo{title}{Conditional image synthesis with auxiliary classifier
  gans}, in: \bibinfo{booktitle}{International conference on machine learning},
  \bibinfo{organization}{PMLR}. pp. \bibinfo{pages}{2642--2651}.
\bibitem[{Oulbacha and Kadoury(2020)}]{oulbacha2020mri}
\bibinfo{author}{Oulbacha, R.}, \bibinfo{author}{Kadoury, S.},
  \bibinfo{year}{2020}.
\newblock \bibinfo{title}{Mri to ct synthesis of the lumbar spine from a
  pseudo-3d cycle gan}, in: \bibinfo{booktitle}{2020 IEEE 17th international
  symposium on biomedical imaging (ISBI)}, pp. \bibinfo{pages}{1784--1787}.
\bibitem[{Pengjiang et~al.(2020)Pengjiang, Xu, Wang, Qiankun, Yang, Atallah,
  Junqing, Bryan and F~Jr}]{Qian2020Estimating}
\bibinfo{author}{Pengjiang, Q.}, \bibinfo{author}{Xu, K.},
  \bibinfo{author}{Wang, T.}, \bibinfo{author}{Qiankun, Z.},
  \bibinfo{author}{Yang, H.}, \bibinfo{author}{Atallah, B.},
  \bibinfo{author}{Junqing, Z.}, \bibinfo{author}{Bryan, T.},
  \bibinfo{author}{F~Jr, M.R.}, \bibinfo{year}{2020}.
\newblock \bibinfo{title}{Estimating ct from mr abdominal images using novel
  generative adversarial networks}.
\newblock \bibinfo{journal}{Journal of Grid Computing} \bibinfo{volume}{18},
  \bibinfo{pages}{211--226}.
\bibitem[{Pourpanah et~al.(2020)Pourpanah, Abdar, Luo, Zhou, Wang, Lim and
  Wang}]{pourpanah2020review}
\bibinfo{author}{Pourpanah, F.}, \bibinfo{author}{Abdar, M.},
  \bibinfo{author}{Luo, Y.}, \bibinfo{author}{Zhou, X.}, \bibinfo{author}{Wang,
  R.}, \bibinfo{author}{Lim, C.P.}, \bibinfo{author}{Wang, X.Z.},
  \bibinfo{year}{2020}.
\newblock \bibinfo{title}{A review of generalized zero-shot learning methods}.
\newblock \bibinfo{journal}{arXiv:2011.08641} .
\bibitem[{Roy et~al.(2014)Roy, Wang, Carass, Prince, Butman and
  Pham}]{roy2014pet}
\bibinfo{author}{Roy, S.}, \bibinfo{author}{Wang, W.T.},
  \bibinfo{author}{Carass, A.}, \bibinfo{author}{Prince, J.L.},
  \bibinfo{author}{Butman, J.A.}, \bibinfo{author}{Pham, D.L.},
  \bibinfo{year}{2014}.
\newblock \bibinfo{title}{Pet attenuation correction using synthetic ct from
  ultrashort echo-time mr imaging}.
\newblock \bibinfo{journal}{Journal of Nuclear Medicine} \bibinfo{volume}{55},
  \bibinfo{pages}{2071--2077}.
\bibitem[{Sangari and Sethares(2015)}]{sangari2015convergence}
\bibinfo{author}{Sangari, A.}, \bibinfo{author}{Sethares, W.},
  \bibinfo{year}{2015}.
\newblock \bibinfo{title}{Convergence analysis of two loss functions in
  soft-max regression}.
\newblock \bibinfo{journal}{IEEE Transactions on Signal Processing}
  \bibinfo{volume}{64}, \bibinfo{pages}{1280--1288}.
\bibitem[{Santini et~al.(2020)Santini, Fourcade, Moreau, Rousseau, Ferrer,
  Lacombe, Fleury, Campone, Jézéquel and Rubeaux}]{santini2020unpaired}
\bibinfo{author}{Santini, G.}, \bibinfo{author}{Fourcade, C.},
  \bibinfo{author}{Moreau, N.}, \bibinfo{author}{Rousseau, C.},
  \bibinfo{author}{Ferrer, L.}, \bibinfo{author}{Lacombe, M.},
  \bibinfo{author}{Fleury, V.}, \bibinfo{author}{Campone, M.},
  \bibinfo{author}{Jézéquel, P.}, \bibinfo{author}{Rubeaux, M.},
  \bibinfo{year}{2020}.
\newblock \bibinfo{title}{Unpaired pet/ct image synthesis of liver region using
  cyclegan}, in: \bibinfo{booktitle}{16th International Symposium on Medical
  Information Processing and Analysis}, \bibinfo{organization}{SPIE}. pp.
  \bibinfo{pages}{247--257}.
\bibitem[{Snell et~al.(2017)Snell, Ridgeway, Liao, Roads, Mozer and
  Zemel}]{snell2017learning}
\bibinfo{author}{Snell, J.}, \bibinfo{author}{Ridgeway, K.},
  \bibinfo{author}{Liao, R.}, \bibinfo{author}{Roads, B.D.},
  \bibinfo{author}{Mozer, M.C.}, \bibinfo{author}{Zemel, R.S.},
  \bibinfo{year}{2017}.
\newblock \bibinfo{title}{Learning to generate images with perceptual
  similarity metrics}, in: \bibinfo{booktitle}{2017 IEEE International
  Conference on Image Processing (ICIP)}, \bibinfo{organization}{IEEE}. pp.
  \bibinfo{pages}{4277--4281}.
\bibitem[{Sohail et~al.(2019)Sohail, Riaz, Wu, Long and
  Li}]{sohail2019unpaired}
\bibinfo{author}{Sohail, M.}, \bibinfo{author}{Riaz, M.N.},
  \bibinfo{author}{Wu, J.}, \bibinfo{author}{Long, C.}, \bibinfo{author}{Li,
  S.}, \bibinfo{year}{2019}.
\newblock \bibinfo{title}{Unpaired multi-contrast mr image synthesis using
  generative adversarial networks}, in: \bibinfo{booktitle}{International
  Workshop on Simulation and Synthesis in Medical Imaging},
  \bibinfo{organization}{Springer}. pp. \bibinfo{pages}{22--31}.
\bibitem[{Ulyanov et~al.(2016)Ulyanov, Vedaldi and
  Lempitsky}]{ulyanov2016instance}
\bibinfo{author}{Ulyanov, D.}, \bibinfo{author}{Vedaldi, A.},
  \bibinfo{author}{Lempitsky, V.}, \bibinfo{year}{2016}.
\newblock \bibinfo{title}{Instance normalization: The missing ingredient for
  fast stylization}.
\newblock \bibinfo{journal}{arXiv preprint arXiv:1607.08022} .
\bibitem[{Wang et~al.(2019)Wang, Manohar, Lei, Dhabaan, Shu, Liu, Curran and
  Yang}]{wang2019mri}
\bibinfo{author}{Wang, T.}, \bibinfo{author}{Manohar, N.},
  \bibinfo{author}{Lei, Y.}, \bibinfo{author}{Dhabaan, A.},
  \bibinfo{author}{Shu, H.K.}, \bibinfo{author}{Liu, T.},
  \bibinfo{author}{Curran, W.J.}, \bibinfo{author}{Yang, X.},
  \bibinfo{year}{2019}.
\newblock \bibinfo{title}{Mri-based treatment planning for brain stereotactic
  radiosurgery: dosimetric validation of a learning-based pseudo-ct generation
  method}.
\newblock \bibinfo{journal}{Medical Dosimetry} \bibinfo{volume}{44},
  \bibinfo{pages}{199--204}.
\bibitem[{Wang et~al.(2020)Wang, Zhao and Pourpanah}]{wang2020recent}
\bibinfo{author}{Wang, X.}, \bibinfo{author}{Zhao, Y.},
  \bibinfo{author}{Pourpanah, F.}, \bibinfo{year}{2020}.
\newblock \bibinfo{title}{Recent advances in deep learning}.
\bibitem[{Wang et~al.(2004)Wang, Bovik, Sheikh and Simoncelli}]{Zhou2004Image}
\bibinfo{author}{Wang, Z.}, \bibinfo{author}{Bovik, A.C.},
  \bibinfo{author}{Sheikh, H.R.}, \bibinfo{author}{Simoncelli, E.P.},
  \bibinfo{year}{2004}.
\newblock \bibinfo{title}{Image quality assessment: from error visibility to
  structural similarity}.
\newblock \bibinfo{journal}{IEEE transactions on image processing}
  \bibinfo{volume}{13}, \bibinfo{pages}{600--612}.
\bibitem[{Wolterink et~al.(2017)Wolterink, Dinkla, Savenije, Seevinck, van~den
  Berg and I{\v{s}}gum}]{wolterink2017deep}
\bibinfo{author}{Wolterink, J.M.}, \bibinfo{author}{Dinkla, A.M.},
  \bibinfo{author}{Savenije, M.H.}, \bibinfo{author}{Seevinck, P.R.},
  \bibinfo{author}{van~den Berg, C.A.}, \bibinfo{author}{I{\v{s}}gum, I.},
  \bibinfo{year}{2017}.
\newblock \bibinfo{title}{Deep mr to ct synthesis using unpaired data}, in:
  \bibinfo{booktitle}{International workshop on simulation and synthesis in
  medical imaging}, pp. \bibinfo{pages}{14--23}.
\bibitem[{Xie et~al.(2022)Xie, Wang, Huang, Zheng, Zheng and
  Jin}]{xie2022fedmed}
\bibinfo{author}{Xie, G.}, \bibinfo{author}{Wang, J.}, \bibinfo{author}{Huang,
  Y.}, \bibinfo{author}{Zheng, Y.}, \bibinfo{author}{Zheng, F.},
  \bibinfo{author}{Jin, Y.}, \bibinfo{year}{2022}.
\newblock \bibinfo{title}{Fedmed-atl: Misaligned unpaired brain image synthesis
  via affine transform loss}.
\newblock \bibinfo{journal}{arXiv preprint arXiv:2201.12589} .
\bibitem[{Xu et~al.(2020)Xu, Zeng, Zhang, Li, Lei and Huang}]{xu2020bpgan}
\bibinfo{author}{Xu, L.}, \bibinfo{author}{Zeng, X.}, \bibinfo{author}{Zhang,
  H.}, \bibinfo{author}{Li, W.}, \bibinfo{author}{Lei, J.},
  \bibinfo{author}{Huang, Z.}, \bibinfo{year}{2020}.
\newblock \bibinfo{title}{Bpgan: Bidirectional ct-to-mri prediction using
  multi-generative multi-adversarial nets with spectral normalization and
  localization}.
\newblock \bibinfo{journal}{Neural Networks} \bibinfo{volume}{128},
  \bibinfo{pages}{82--96}.
\bibitem[{Yang et~al.(2020)Yang, Sun, Carass, Zhao, Lee, Prince and
  Xu}]{Heran2020Unsupervised}
\bibinfo{author}{Yang, H.}, \bibinfo{author}{Sun, J.}, \bibinfo{author}{Carass,
  A.}, \bibinfo{author}{Zhao, C.}, \bibinfo{author}{Lee, J.},
  \bibinfo{author}{Prince, J.L.}, \bibinfo{author}{Xu, Z.},
  \bibinfo{year}{2020}.
\newblock \bibinfo{title}{Unsupervised mr-to-ct synthesis using
  structure-constrained cyclegan}.
\newblock \bibinfo{journal}{IEEE transactions on medical imaging}
  \bibinfo{volume}{39}, \bibinfo{pages}{4249--4261}.
\bibitem[{Yang et~al.(2021)Yang, Kim and Ye}]{yang2021continuous}
\bibinfo{author}{Yang, S.}, \bibinfo{author}{Kim, E.Y.}, \bibinfo{author}{Ye,
  J.C.}, \bibinfo{year}{2021}.
\newblock \bibinfo{title}{Continuous conversion of ct kernel using switchable
  cyclegan with adain}.
\newblock \bibinfo{journal}{IEEE Transactions on Medical Imaging}
  \bibinfo{volume}{40}, \bibinfo{pages}{3015--3029}.
\bibitem[{Yang et~al.(2018)Yang, Zhong, Chen, Lin, Lu, Liu, Wu, Feng and
  Chen}]{yang2018predicting}
\bibinfo{author}{Yang, W.}, \bibinfo{author}{Zhong, L.}, \bibinfo{author}{Chen,
  Y.}, \bibinfo{author}{Lin, L.}, \bibinfo{author}{Lu, Z.},
  \bibinfo{author}{Liu, S.}, \bibinfo{author}{Wu, Y.}, \bibinfo{author}{Feng,
  Q.}, \bibinfo{author}{Chen, W.}, \bibinfo{year}{2018}.
\newblock \bibinfo{title}{Predicting ct image from mri data through feature
  matching with learned nonlinear local descriptors}.
\newblock \bibinfo{journal}{IEEE transactions on medical imaging}
  \bibinfo{volume}{37}, \bibinfo{pages}{977--987}.
\bibitem[{Yi et~al.(2017)Yi, Zhang, Tan and Gong}]{yi2017dualgan}
\bibinfo{author}{Yi, Z.}, \bibinfo{author}{Zhang, H.}, \bibinfo{author}{Tan,
  P.}, \bibinfo{author}{Gong, M.}, \bibinfo{year}{2017}.
\newblock \bibinfo{title}{Dualgan: Unsupervised dual learning for
  image-to-image translation}, in: \bibinfo{booktitle}{Proceedings of the IEEE
  international conference on computer vision}, pp.
  \bibinfo{pages}{2849--2857}.
\bibitem[{Yu et~al.(2019)Yu, Zhou, Wang, Shi, Fripp and Bourgeat}]{yu2019ea}
\bibinfo{author}{Yu, B.}, \bibinfo{author}{Zhou, L.}, \bibinfo{author}{Wang,
  L.}, \bibinfo{author}{Shi, Y.}, \bibinfo{author}{Fripp, J.},
  \bibinfo{author}{Bourgeat, P.}, \bibinfo{year}{2019}.
\newblock \bibinfo{title}{Ea-gans: edge-aware generative adversarial networks
  for cross-modality mr image synthesis}.
\newblock \bibinfo{journal}{IEEE transactions on medical imaging}
  \bibinfo{volume}{38}, \bibinfo{pages}{1750--1762}.
\bibitem[{Yueyun and Junping(2008)}]{Yueyun2008ANALYSIS}
\bibinfo{author}{Yueyun, B.}, \bibinfo{author}{Junping, L.},
  \bibinfo{year}{2008}.
\newblock \bibinfo{title}{Analysis of the advantages and disadvantages of ct,
  mri and b ultrasound for their reasonable use}.
\newblock \bibinfo{journal}{Modern Hospital} , \bibinfo{pages}{62--63}.
\bibitem[{Zeng and Zheng(2019)}]{zeng2019hybrid}
\bibinfo{author}{Zeng, G.}, \bibinfo{author}{Zheng, G.}, \bibinfo{year}{2019}.
\newblock \bibinfo{title}{Hybrid generative adversarial networks for deep mr to
  ct synthesis using unpaired data}, in: \bibinfo{booktitle}{International
  Conference on Medical Image Computing and Computer-Assisted Intervention},
  pp. \bibinfo{pages}{759--767}.
\bibitem[{Zhang et~al.(2018)Zhang, Yang and Zheng}]{zhang2018translating}
\bibinfo{author}{Zhang, Z.}, \bibinfo{author}{Yang, L.},
  \bibinfo{author}{Zheng, Y.}, \bibinfo{year}{2018}.
\newblock \bibinfo{title}{Translating and segmenting multimodal medical volumes
  with cycle-and shape-consistency generative adversarial network}, in:
  \bibinfo{booktitle}{Proceedings of the IEEE conference on computer vision and
  pattern Recognition}, pp. \bibinfo{pages}{9242--9251}.
\bibitem[{Zhou et~al.(2021)Zhou, Liu, Pourpanah, Zeng and
  Wang}]{zhou2021survey}
\bibinfo{author}{Zhou, X.}, \bibinfo{author}{Liu, H.},
  \bibinfo{author}{Pourpanah, F.}, \bibinfo{author}{Zeng, T.},
  \bibinfo{author}{Wang, X.}, \bibinfo{year}{2021}.
\newblock \bibinfo{title}{A survey on epistemic (model) uncertainty in
  supervised learning: Recent advances and applications}.
\newblock \bibinfo{journal}{Neurocomputing} .
\bibitem[{Zhou et~al.(2019)Zhou, Wang, Zhang, Zhu, Zheng and Wu}]{zhou2019mpce}
\bibinfo{author}{Zhou, Y.}, \bibinfo{author}{Wang, X.}, \bibinfo{author}{Zhang,
  M.}, \bibinfo{author}{Zhu, J.}, \bibinfo{author}{Zheng, R.},
  \bibinfo{author}{Wu, Q.}, \bibinfo{year}{2019}.
\newblock \bibinfo{title}{Mpce: a maximum probability based cross entropy loss
  function for neural network classification}.
\newblock \bibinfo{journal}{IEEE Access} \bibinfo{volume}{7},
  \bibinfo{pages}{146331--146341}.
\bibitem[{Zhu et~al.(2017)Zhu, Park, Isola and Efros}]{zhu2017unpaired}
\bibinfo{author}{Zhu, J.Y.}, \bibinfo{author}{Park, T.},
  \bibinfo{author}{Isola, P.}, \bibinfo{author}{Efros, A.A.},
  \bibinfo{year}{2017}.
\newblock \bibinfo{title}{Unpaired image-to-image translation using
  cycle-consistent adversarial networks}, in: \bibinfo{booktitle}{Proceedings
  of the IEEE international conference on computer vision}, pp.
  \bibinfo{pages}{2223--2232}.

\end{thebibliography}





\end{document}